\documentclass[aps,pra,twocolumn,reprint,groupedaddress]{revtex4-1}

\usepackage{physics}
\usepackage[caption=false,subrefformat=parens,labelformat=parens]{subfig}
\usepackage{graphicx, color, graphpap}
\usepackage{amsmath}
\usepackage{nccmath}
\usepackage{enumitem}
\usepackage{amssymb}
\usepackage{bm}
\usepackage{dsfont}
\usepackage{mathrsfs}
\DeclareMathOperator{\spn}{span}
\renewcommand{\vec}[1]{\boldsymbol{#1}}
\usepackage{hyperref}

\begin{document}

\title{Simple security analysis of phase-matching measurement-device-independent \\ quantum key distribution}

\author{Jie Lin}
\affiliation{Institute for Quantum Computing and Department of Physics and Astronomy, University of Waterloo, Waterloo, Ontario, Canada N2L 3G1} 
\author{Norbert L\"utkenhaus}
\affiliation{Institute for Quantum Computing and Department of Physics and Astronomy, University of Waterloo, Waterloo, Ontario, Canada N2L 3G1}

\date{\today}

\begin{abstract}
Variations of phase-matching measurement-device-independent quantum key distribution (PM-MDI QKD) protocols have been investigated before, but it was recently discovered that this type of protocol (under the name of twin-field QKD) can beat the linear scaling of the repeaterless bound on secret key capacity. We propose a variation of PM-MDI QKD protocol, which reduces the sifting cost and uses non-phase-randomized coherent states as test states. We provide a security proof in the infinite key limit. Our proof is conceptually simple and gives tight key rates. We obtain an analytical key rate formula for the loss-only scenario, confirming the square root scaling and also showing the loss limit. We simulate the key rate for realistic imperfections and show that PM-MDI QKD can overcome the repeaterless bound with currently available technology.

\end{abstract}

\maketitle

\section{INTRODUCTION}
Quantum key distribution (QKD) \cite{Bennett1984, Ekert1991} protocols enable two distant parties (Alice and Bob) to establish information-theoretically secure private keys using a quantum channel and an authenticated classical channel. There is a wealth of QKD protocols around (see Ref. \cite{Scarani2009} for a review). A bottleneck for QKD applications, be it as individual link or as part of a network, is the scaling of the generated secret key rate with the loss in the channel represented by the single-photon transmissivity $\eta$. The best-known QKD protocols have a scaling of their key rate in the limit of infinite channel uses (infinite key limit) as $R^\infty = O(\eta)$, and by now we have bounds on repeaterless optical channels which show that this is the optimal rate scaling that can be achieved \cite{Takeoka2014,Pirandola2017}. The tight bound on the performance of QKD in terms of secret key rate per employed optical mode is given by $R^\infty \leq \log_2 \frac{1}{1-\eta}$  \cite{Pirandola2017}, which can be saturated \cite{Pirandola2009}. In principle, inserting intermediate stations performing some operations can improve the performance, and quantum repeaters \cite{Briegel1998} aim at this. The field of quantum repeater research is very active and made conceptual and practical advances over the recent years, but as of today, no quantum repeater has been demonstrated yet that would outperform the direct use of optical channels, and thus breaking the repeaterless bounds.

While proposals have been made for simplest possible devices that allow a demonstration of quantum repeater action by beating repeaterless bounds using a simple single-node layout \cite{Luong2016}, the corresponding quantum advantage has not been experimentally demonstrated yet. In a pleasant surprise to the field, the phase-matching measurement-device-independent protocols (PM-MDI) \cite{Tamaki2012,Ferenczi2013} were recently shown to beat the repeaterless bound when using suitable test states \cite{Lucamarini2018}. This important observation justifiably creates quite an interest in the community. In the original paper \cite{Lucamarini2018}, it has been argued that the secret key rate in the infinite key limit indeed scales as $R^{\infty}= O(\sqrt{\eta})$, where we keep $\eta$ as the single-photon transmissivity of the total distance, rather than that of a segment. It is interesting to see that an MDI protocol can achieve that performance without the use of any quantum memory or similar advanced components. Remarkably, the only difference to previous MDI QKD protocols that show a scaling of $R^\infty=O(\eta)$ is the change from single-photon signals (or mixture of photon number states) with two-photon interference events at the beam splitter, to coherent states as signal states and single-photon interference events at the beam splitter. 

So far, the security analyses \cite{Ma2018,Tamaki2018} of the PM-MDI QKD protocols have been done in a framework based on the quantum error correction inspired approach by Shor-Preskill \cite{Shor2000}, which is improved by Koashi \cite{Koashi2009}, and later extended to work with a wider class of privacy amplification protocols \cite{Tsurumaru2013}. However, for security proofs in this framework, due to some pessimistic estimation of the phase error rate, the key rate bound can potentially be loose. Also, the variations of the protocol proposed in \cite{Ma2018,Tamaki2018} require phase-randomization on the signal states and thus introduce a large sifting cost. 
The goal of the present paper is two-fold. We propose a variation of the PM-MDI QKD protocol that clearly distinguishes between test states, meant to probe potential eavesdropping activities, and signal states, which are meant to establish secret keys and are not phase randomized. For this modified protocol we then execute a security analysis which is expected to be tight as it uses the framework by Renner \cite{Renner2005}. This framework is known to be flexible in terms of error correction and privacy amplification methods, and is general enough to be adaptable to any generic QKD protocol. It is interesting to point out that even though these two frameworks have been generally considered independent of each other in the community, recently there has been an effort to unify these two different security proof frameworks \cite{Tsurumaru2018}.

We will first analyze the security of the protocol in a setting with infinitely many different test states, similar to the initial discussion of decoy states in weak coherent pulse BB84 protocols \cite{Lo2005}. In this setting, we can derive an analytical key rate formula for the scenario where Alice and Bob observe correlations coming from a loss-only scenario. We derive the general framework that includes also the noisy case, for which we then resort to numerical evaluations to demonstrate the stability of the proposed protocol. 

Since the test states in our protocol are non-phase-randomized coherent states, we do not apply the decoy state analysis. Instead of using phase-randomized coherent states to simulate a classical mixture of photon number states, we directly use the properties of coherent states to perform a variation of tomography on the quantum channel and untrusted measurement devices. Our approach thus generalizes the decoy state idea to the use of general test states to test the channel and to deduce information about the adversary's attacks. To the best of our knowledge, this approach has not been used in the context of QKD security proof and might be interesting to apply a similar approach to other QKD protocols.  

This paper is organized as follows. We first describe our version of the PM-MDI protocol in Sec. \ref{sec:description}, and we then compare different variations of the PM-MDI protocol in Sec. \ref{sec:comparison} and highlight how our variation (and our proof idea) differs from other works. Next, we describe the framework for our security proof and procedures for key rate calculation in Sec. \ref{sec:security}. We then simulate the key rates with the loss-only scenario and with realistic experimental imperfections in Sec. \ref{sec:simulation}. Finally, in Sec. \ref{sec:outlook}, we summarize our results and provide insights for future work. Some technical details relevant for the key rate calculation are presented in the appendixes.



\section{PM-MDI QKD PROTOCOLS}
In this section, we first present an idealized version of PM-MDI QKD in the sense that Alice and Bob use infinitely many coherent states as test states in the protocol, similar to the initial discussion of decoy states in weak coherent pulse BB84 protocols \cite{Lo2005}. We will prove its security in this paper. Then, we will compare different variations of PM-MDI QKD protocols. In the next section, when we prove the security of the idealized version of our protocol, we will also provide insights for the security analysis of a practical version of this protocol with a small number of choices of test states.

\subsection{Description of our protocol}\label{sec:description}

\begin{figure}
\subfloat[]{\label{fig:scheme_a}\includegraphics[width=0.48\linewidth]{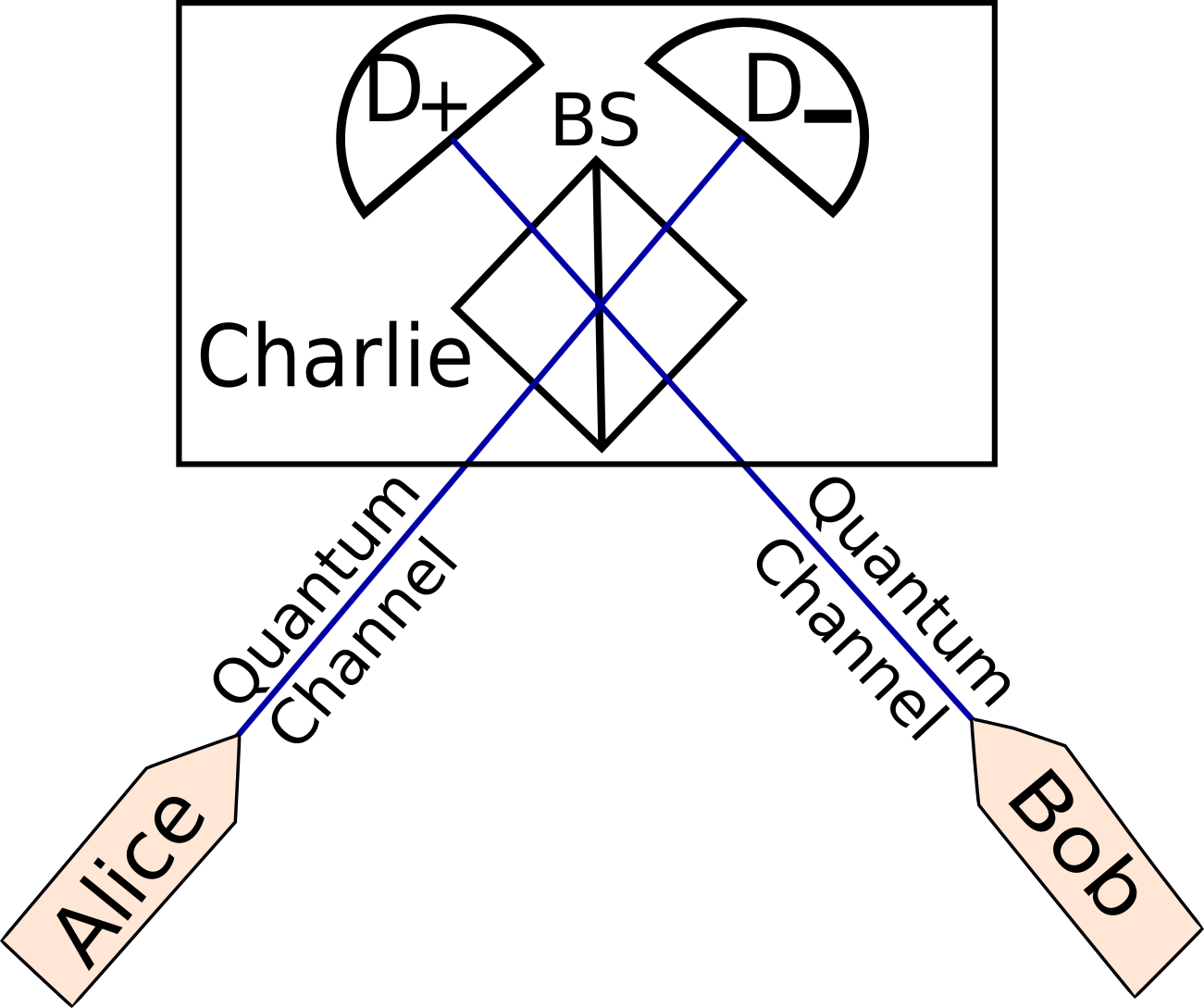}}
\subfloat[]{\label{fig:scheme_b}\includegraphics[width=0.48\linewidth]{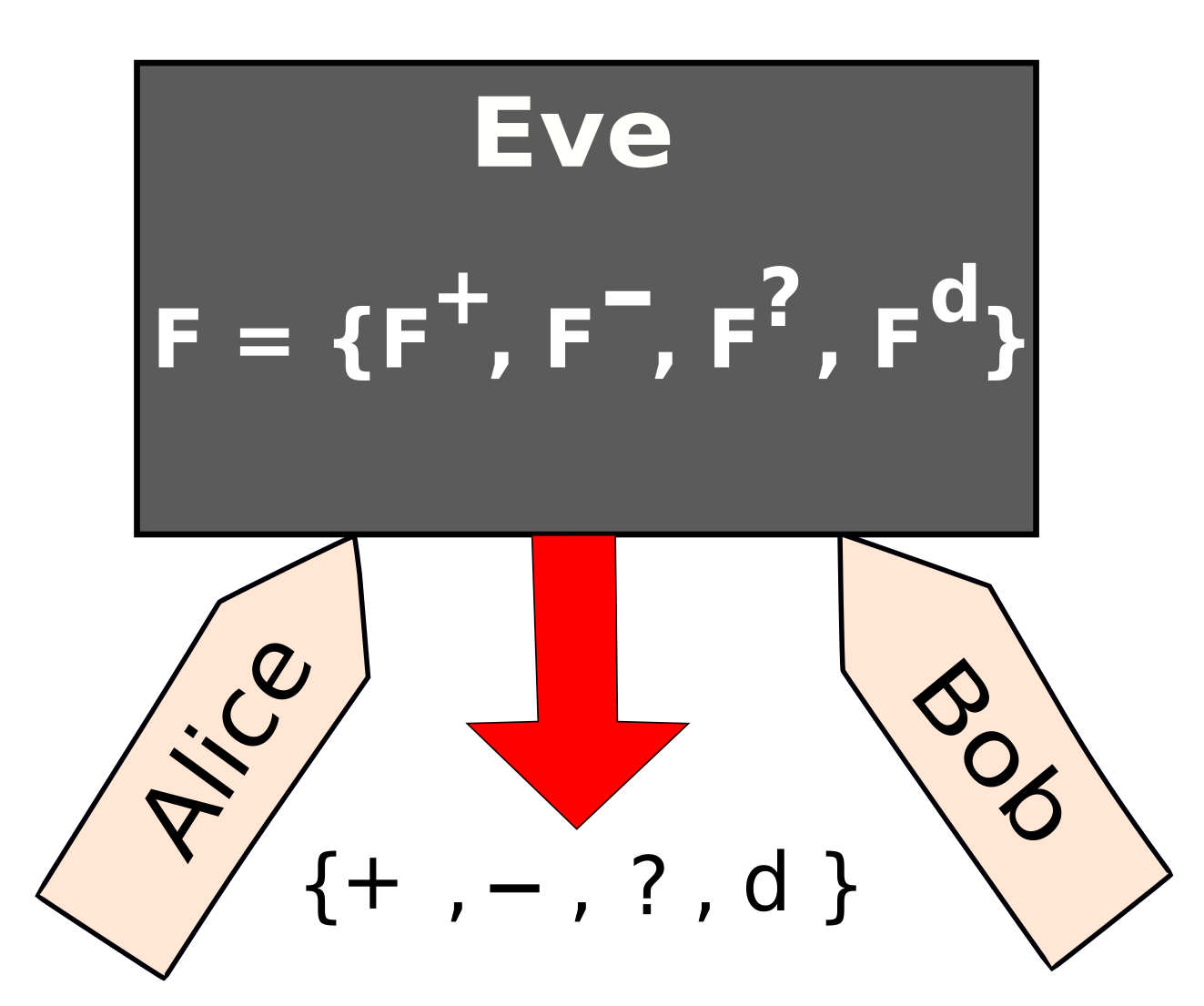}}
\caption{\label{fig:scheme}  (a). Schematic setup of the PM-MDI QKD protocol. Alice and Bob send coherent states to the untrusted third party Charlie in the middle, who performs measurements and broadcasts outcomes. BS: 50-50 beam splitter. $D_+$, $D_-$: single-photon detectors. (b). Equivalent view of the protocol. Eve is assumed to perform the measurements in the middle. Effectively, Eve performs a 4-element POVM, denoted as $\{F^+, F^-, F^?, F^d\}$, corresponding to four possible announcements $\{D_+$ clicks, $D_-$ clicks, no detectors click, both detectors click$\},$ which are abbreviated as $\{+,-,?,d\}.$ }
 \end{figure}

\begin{enumerate}[leftmargin=*,wide]
\item[(1)] Test/key-generation mode selection. Alice (Bob) chooses a random bit $m_A$ ($m_B$) according to \textit{a priori} probability distribution $\{p_A, 1-p_A\}$ ($\{p_B, 1-p_B\}$). If $m_A = 0$, Alice then labels this round as in the key-generation mode. If $m_A =1$, Alice labels this round as in the test mode and similarly for Bob. 
\item[(2)] State preparation. If the test mode is chosen,  Alice (Bob) then randomly chooses a phase $\theta_A$ ($\theta_B$) $\in [0, 2\pi)$ and randomly chooses an intensity $\mu_A$ ($\mu_B$). Then she (he) prepares a coherent state $\ket{\sqrt{\mu_A} e^{i  \theta_{A}}}$ ( $\ket{\sqrt{\mu_B} e^{i \theta_{B}}}$) and sends it to the untrusted third party Charlie through the quantum channel.

If the key-generation mode is chosen, Alice (Bob) randomly generates a bit value $k_A$  ($k_B$) $\in \{0,1\}$ with a uniform probability distribution. Alice (Bob) chooses the pre-agreed intensity $\mu$ and sends a coherent state $\ket{\sqrt{\mu} e^{i \pi k_A}}$ ($\ket{\sqrt{\mu} e^{i \pi k_B}}$) to Charlie.
\item[(3)] Measurements. For each round, Charlie performs a joint measurement on the signals received from Alice and Bob, and then makes an announcement about the measurement outcome. If Charlie is honest, he is supposed to perform the measurement as shown in Fig. \subref*{fig:scheme_a} and announces one of the following outcomes $\{$``Only detector $D_+$ clicks", ``Only detector  $D_-$  clicks", ``No detectors click", ``Both detectors click"$\}$, which, for the later convenience of notation, we abbreviate as $\{+,-,?,d\}$, respectively. We denote Charlie's announcement as $\gamma$ throughout this paper.
\end{enumerate}
After steps 1-3 are repeated for many times, and after Charlie has made all the announcements, Alice and Bob then proceed with the following steps.
\begin{enumerate}[leftmargin=*, wide]
\item[(4)] Sifting. Alice and Bob use an authenticated classical channel to communicate and sort all rounds into two disjoint sets, where one set is used for the key generation and the other is for the parameter estimation. To do so, they disclose the choices of $m_A$ and $m_B$ for each round and also use the announcement $\gamma$. If $m_A = m_B =0$, that is, they both selected the key-generation mode for a given round, and Charlie announced $\gamma \in \{+,-\}$, they save their data corresponding to this round for the key generation. All remaining rounds are used for parameter estimation.

\item[(5)] Parameter estimation. To perform parameter estimation, Alice and Bob disclose the choices of $\mu_A, \mu_B, \theta_A, \theta_B$ (also $k_A, k_B$ if they have chosen one for that round) for the rounds in the set labeled for parameter estimation and also use the announcement result $\gamma$ for each of these rounds to estimate how Eve has interacted with the signals during their exchange in the protocol. If, from their analysis, they find out that Eve has learned too much about the signals and no secret keys can be generated, then they abort the protocol. Otherwise, they continue. 

\item[(6)] Key map. Alice forms a raw key using her bit value $k_A$ from each of the rounds saved for the key generation. (In principle, Bob does not need to do anything in this step since he can correctly determine Alice's key by the error correction. In practice, depending on the choice of error correction code, it might be convenient for Bob to flip his bit value when the announcement is $\gamma = ``-"$.)

\item[(7)] Error correction and privacy amplification. Alice and Bob then apply the procedures of error correction and privacy amplification as in a typical QKD protocol to generate a secret key.
\end{enumerate}

We remark that since this protocol uses an MDI setup, it is inherently immune to all side channels in the measurement devices once its security is proven. However, Alice and Bob's sources have to be trusted and protected. In our security analysis, we assume that Alice and Bob's devices are fully characterized and Eve has no access. This assumption needs to be justified in the experimental implementations of the protocol. In particular, we want to remark that the choices of $m_A$ and $m_B$ (also $k_A$ and $k_B$) should not be leaked to Eve by side channels before the announcement is made. In the implementation of the protocol, Alice and Bob need to make sure that Eve cannot distinguish the key-generation mode from test mode by any classical side information leaked from their devices before Charlie's announcement. Just like other MDI QKD protocols, this protocol can be vulnerable to side-channel attacks on the sources.   

We also comment on the the choices of parameters $p_{A}$, $p_{B}$. While values of $p_{A}$, $p_{B}$ need to be optimized in the finite-key regime, in the infinite key limit, we can choose $p_{A}$ and $p_{B}$ arbitrarily close to 1 so that the sifting factor is asymptotically 1, like the efficient BB84 protocol \cite{Lo2004}. 

Finally, we remark on the choices of $\mu_A$ and $\mu_B$ and their corresponding probability distributions. Since states in the test mode essentially are used to perform a tomography on Eve's attacks on the subspace of signal states used in the key-generation mode, for the purpose of this paper, we initially use coherent states whose complex amplitudes cover the entire complex space. In the infinite key limit, the probability distribution (with no zeros) does not matter. We will remark on how a finite number of choices of test states can approximately accomplish the same task and the choices of  $\mu_A$ and $\mu_B$ will then be closely related to the value of $\mu$.

\subsection{Comparison of different variations of PM-MDI QKD protocols}\label{sec:comparison}
Many variations of the PM-MDI QKD protocol have been proposed and investigated. Different names have been assigned to different variations, such as, phase-encoding scheme for MDI \cite{Tamaki2012}, MDI-B92 \cite{Ferenczi2013}, twin-field QKD (TF-QKD) \cite{Lucamarini2018} and phase-matching QKD (PM-QKD) \cite{Ma2018}. However, they all have the essential components needed to achieve the rate scaling of $R^{\infty} = O(\sqrt{\eta})$, namely, they all use coherent states as signal states and rely on single-photon interference events at the beam splitter of an untrusted intermediate node, even though not all variations can indeed achieve this scaling.    

We first describe the common features of all those protocols and then discuss how each variation differs in the following two aspects: choices of signal states used for establishing secret keys and choices of test states used to probe Eve's attacks. 

In an ideal PM-MDI QKD protocol, Alice and Bob will only establish keys from the  rounds where each of them has selected a state from the set $\{\ket{+\alpha}, \ket{-\alpha}\}$, where $\alpha$ can be an arbitrary complex number. In other words, Alice and Bob will only establish keys from the rounds that satisfies the phase-matching condition, that is, they have chosen the same global phase and same intensity for their states. We call two coherent states $\{\ket{+\alpha}, \ket{-\alpha}\}$ with only a $\pi$ phase difference as a phase-matching pair. In addition, Alice and Bob may decide to send some states as test states to probe eavesdropping activities for randomly selected rounds and those rounds will be used in the parameter estimation step only. Alice and Bob will send their states to an untrusted party Charlie at the intermediate node for measurements. An honest Charlie will use the single-photon interference events at the beam splitter for his announcement.  

Since this type of protocol is measurement-device independent and generates keys when Alice and Bob use the same phase-matching pair of coherent states, phase-matching measurement-device-independent QKD is in our view a more descriptive name that captures important features of this type of protocol. 

Now, we compare some variations of PM-MDI QKD. Different variations may use different number of phase-matching pairs as signal states and may use different types of states as test states, such as a mixture of photon number states (phase-randomized coherent states), partially phase-randomized coherent states, or coherent states without phase randomization. Some variations may use the same number of phase-matching pairs as signal states, but differ in how to handle them. We present those variations just for the comparison purpose and we do not neither claim this is an exhausted list nor verify the security analysis of each work.
\begin{enumerate}[leftmargin=*, wide]

\item[(1)] The variation proposed in Ref. \cite{Tamaki2012} is called phase encoding scheme I for MDI. This protocol essentially uses two phase-matching pairs of coherent states. In the original description of the protocol, these two pairs are labeled as two bases, similar to a BB84-type protocol, due to the proof technique adopted. In an abstract description, we can view this protocol as essentially using one phase-matching pair of coherent states as signal states and an additional pair as test states. Because of the proof technique and a limited number of test states, the scaling $R^{\infty} = O(\sqrt{\eta})$ was not found. 
\item[(2)] The variation studied in Ref. \cite{Ferenczi2013} is called MDI-B92 protocol. Reference \cite{Ferenczi2013} analyzes different types of measurements for the intermediate node. Under the investigation of unambiguous state discrimination attacks, it basically proposes a variation of PM-MDI protocol with exactly one phase-matching pair of coherent states as the signal states and no test states. Because there are no test states, this protocol is not expected to have the scaling $R^{\infty} = O(\sqrt{\eta})$.
\item[(3)]  The variation proposed in Ref. \cite{Lucamarini2018} has the name of TF-QKD protocol. This protocol uses infinitely many phase-matching pairs of coherent states (phase-randomized coherent states) as signal states. In addition, for the purpose of security analysis, each round is assigned to one of two bases to mimic a BB84-type protocol. Instead of achieving the perfect phase-matching conditions, this protocol allows some small errors in identifying whether Alice and Bob have chosen the same phase-matching pair. To distill keys, Alice and Bob disclose some partial information about the global phases. If their global phases only differ by a small amount, they assume they have used the same phase-matching pair. Due to the phase-matching condition, the sifting cost of this protocol can be large, which affects the prefactor of key rate. In this protocol, states used as test states are effectively the same as states used for signal states. These test states are partially phase-randomized coherent states as Eve knows some partial information about the global phase. Reference \cite{Lucamarini2018} argued that this type of protocol can have the $O(\sqrt{\eta})$ rate scaling.
\item[(4)] The variation investigated in Ref. \cite{Ma2018} uses the name PM-QKD protocol. Like TF-QKD \cite{Lucamarini2018}, it also uses infinitely many phase-matching pairs of coherent states as signal states and adopts a procedure similar to that in Ref. \cite{Lucamarini2018} in identifying whether Alice and Bob have chosen the same phase-matching pair for each round. It also uses partially phase-randomized coherent states as test states. The difference from TF-QKD is that there is no assignment of basis choice for each round. This work shows the $O(\sqrt{\eta})$ rate scaling and its security analysis does not use standard decoy state methods.
\item[(5)] The variation studied in Ref. \cite{Tamaki2018} is called TF-QKD$^*$ protocol. This protocol, similar to the original TF-QKD protocol, uses infinitely many phase-matching pairs as signal states and later postselects on rounds where the global phases are different by less than a small amount. Effectively, by allowing some errors, Alice and Bob assume that they have chosen the same phase-matching pair when the difference in their global phases is small. This protocol also has an assignment of basis choice for each round in order to apply a BB84-type security argument. Unlike the original TF-QKD protocol, this protocol uses a mixture of photon number states as test states. The security analysis applies the standard decoy state methods and shows $O(\sqrt{\eta})$ rate scaling.
\item[(6)] The variation proposed in Ref. \cite{Cui2018} is also called PM-QKD protocol. It uses exactly one phase-matching pair as signal states and uses a mixture of photon number states as test states. 
\item[(7)] The variation studied in Ref. \cite{Curty2018} is referred as a TF-QKD type protocol. This variation essentially is the same as in Ref. \cite{Cui2018}. It uses exactly one phase-matching pair as signal states and uses a mixture of photon number states as test states. These two works differ by the security proof methods. 
\item[(8)] In this paper, we propose a modified PM-MDI QKD protocol. Our protocol uses exactly one phase-matching pair as signal states and infinitely many different coherent states (without phase randomization) as test states. Our security analysis does not use the standard decoy state method since our test states are not mixtures of photon number states. We rely on the tomographic reconstruction of POVM elements using the coherent states as test states to prove the security. 

\end{enumerate}

In the end, we remark on the advantages of different types of test states. Phase-randomized weak coherent state sources are used to approximate single-photon sources and test states of this type are used to estimate the single-photon contribution. Using a mixture of photon number states as test states allows the standard decoy state analysis, which has been investigated and well understood. In addition, using a small number of decoy states \cite{Ma2005, Zhou2016} as test states has been investigated in many other protocols and might be readily adapted to some variations of PM-MDI QKD protocol. On the other hand, using non-phase-randomized coherent states as test states, we directly use properties of coherent states, that is, they are overcomplete and non-orthogonal to each other. Test states of this type have the potential to give tighter key rates, as we will demonstrate in this paper when using infinitely many coherent states. Also, no phase randomization is required in the experimental implementations. 

\section{SECURITY PROOF}\label{sec:security}
To prove the security of a QKD protocol, the ultimate goal is to provide a full security proof following the $\epsilon$-security definition of QKD \cite{Renner2005, Muller-Quade2009} in the framework of universal composability. Currently, there are well-developed techniques to simplify the problem, such as, the quantum de Finetti theorem \cite{Renner2005}, so-called postselection technique \cite{Christandl2009}, and the entropy accumulation theorem \cite{Dupuis2016, Dupuis2018}. These techniques allow us to prove the security in a two-step procedure. In the first step, we prove the security against collective attacks in the infinite key scenario, and in the second step we apply one of the mentioned techniques to extend the analysis to a full security proof against general attacks, including finite-size effects. The scope of this paper is to prove the first step, namely the analysis of the collective attack in the infinite key limit. We leave technical details of the extension to the full security for the future work.

To prove the security of this protocol against collective attacks, we first apply the source-replacement scheme \cite{Curty2004, Ferenczi2012} to both Alice and Bob's sources and convert this protocol to its equivalent entanglement-based protocol. Then we proceed to prove the security of the entanglement-based version by evaluating the secret key generation rate.

\subsection{Source-replacement scheme}
For the purpose of clarifying notations, let us start with a more abstract view of the protocol. In each round, Alice chooses a state from the set of possible signal states $\{\ket{\varphi_x}\}$ according to \textit{a priori} probability distribution $\{p_x\}$ and similarly, Bob chooses a state from the same set $\{\ket{\varphi_y}\}$ with \textit{a priori} probability distribution $\{q_y\}$. Then in the source-replacement scheme, Alice and Bob's sources effectively prepare the following state 
\begin{equation}\label{eq:sourcereplacement}
\begin{aligned}
&\ket{\Psi}_{ABA'B'} \\
=&\Big(\sum_{x} \sqrt{p_x} \ket{x}_{A} \ket{\varphi_x}_{A'}\Big) \otimes \Big(\sum_{y} \sqrt{q_y} \ket{y}_{B} \ket{\varphi_y}_{B'}\Big) \\ 
=&\sum_{x,y} \sqrt{p_x q_y} \ket{x,y}_{AB} \ket{\varphi_x, \varphi_y}_{A'B'}, 
\end{aligned}
\end{equation}
where the register $A$ records the choices of states prepared in the register $A'$ and similarly the register $B$ records the choices of states in the register $B'$. We introduce an orthonormal basis $\{\ket{x}_A\}$ for the register system $A$ corresponding to states $\{\ket{\varphi_x}\}$, and an orthonormal basis $\{\ket{y}_B\}$ for the register system $B$ corresponding to states $\{\ket{\varphi_y}\}$. It is crucial that Eve has no access to the registers $A$ and $B$. Then, Alice keeps the register $A$ and sends the system $A'$ to Charlie, and similarly, Bob keeps $B$ and sends $B'$. To learn their choices of states sent to Charlie for each round, Alice performs a local measurement described by a positive-operator valued measure (POVM) $M_A=\{\dyad{x}{x}\}$ on her register $A$ and likewise, Bob applies his POVM $M_B=\{\dyad{y}{y}\}$ to his register $B$.

Importantly, we only apply the source-replacement scheme for the signal states in the key-generation mode since the test states in the test mode are only used to put constraints on how Eve acts in the subspace spanned by signal states. We denote the set of signal states in the key-generation mode as $\mathcal{S}$, that is, 
\begin{equation}\label{eq:definitionS}
\begin{aligned}
\mathcal{S} =& \{\ket{+\sqrt{\mu},+\sqrt{\mu}}, \ \  \ket{-\sqrt{\mu},-\sqrt{\mu}}, \\ &\ \ket{+\sqrt{\mu},-\sqrt{\mu}}, \ \ \ket{-\sqrt{\mu},+\sqrt{\mu}}\},
\end{aligned}
\end{equation}
where each state is a two-mode coherent state coming from both Alice and Bob, and we dropped the subscript $A'B'$ for the ease of writing. Since finitely many coherent states are linearly independent, we want to point out that $\mathcal{S}$ is indeed a basis of $\spn(\mathcal{S})$.

\subsection{Description of Eve's attack}
As an MDI QKD protocol, Eve has a full control of both the quantum channels connecting Alice, Bob and the intermediate node Charlie, and the measurement devices at the intermediate node. Since measurement devices are neither characterized nor trusted, Eve is assumed to play the role of Charlie to perform the measurement. Therefore, in the PM-MDI QKD protocol, we can view the protocol in an alternative and equivalent picture, as shown in Fig. \subref*{fig:scheme_b}. In order to make an announcement strategy, Eve performs some measurement, which can be described by a POVM $F$, directly on the states from Alice and Bob in the registers $A'$ and $B'$. Moreover, without loss of generality, we can assume that $F$ only has four elements since only $\{+,-,?,d\}$ outcomes are meaningful for Alice and Bob, and all other outcomes are simply discarded in the protocol. (Even though Alice and Bob may only keep $\{+,-\}$ outcomes to distill keys, we are allowed to include $\{?, d\}$ outcomes for parameter estimation.)  We write this POVM $F$ as $F=\{F^+, F^-,F^?, F^d\}$, or abbreviate it as $\{F^{\gamma}\}$ for $\gamma \in \{+,-,?,d\}$. The probability of announcing the outcome $\gamma$ is $\Tr(F^{\gamma} \sigma_{A'B'})$ for an input state $\sigma_{A'B'}$.

From Alice and Bob's point of view, they can only know the probability of each announcement, not the post-measurement states in Eve's hand. They can infer what POVM $F$ that Eve applied from their observed correlations. However, Eve can perform a nondestructive measurement and keep her post-measurement states for further analysis. That is, Eve applies a completely positive trace-preserving (CPTP) map $\mathcal{E}_{A'B' \rightarrow EC}$ on the input quantum states in the registers $A'$ and $B'$. Her announcement about the measurement outcome is stored in the classical register $C$ and she keeps the postmeasurement state in the register $E$. Here, we introduce an orthonormal basis $\{\ket{\gamma}\}$ for the register $C$, each of which corresponds to every possible announcement outcome. In general, we can write $\mathcal{E}_{A'B' \rightarrow EC}$ as follows:
\begin{equation}\label{eq:Etot}
\mathcal{E}_{A'B' \rightarrow EC} (X) =\sum_{\gamma} \mathcal{E}_{\gamma} (X) \otimes \dyad{\gamma}{\gamma}_C,
\end{equation}
where each $\mathcal{E}_{\gamma}$ is a completely positive trace non-increasing map and $X$ is an arbitrary linear operator on the systems $A'B'$. 

In the Choi-Kraus representation, each $\mathcal{E}_{\gamma}$ can be written as
\begin{equation}
\label{eq:EGen}
\mathcal{E}_{\gamma}(X) = \sum_{j \in \mathcal{I}(\gamma)} K_j^{\gamma} X (K_j^{\gamma})^{\dagger},
\end{equation}
with $\sum_{j}  (K_j^{\gamma})^{\dagger}K_j^{\gamma} = F^{\gamma}$ and the summation going over some index set $\mathcal{I}(\gamma)$ that depends on $\gamma$.  Without loss of generality we can use maps $\mathcal{E}_{\gamma}(X)$ with a single Kraus operator $K^{\gamma} = \sqrt{F^{\gamma}}$. The reason for this is that the general case of Eqs. (\ref{eq:Etot}) and (\ref{eq:EGen}) can be represented as a concatenation of two maps, the first one using the case of $K^{\gamma} = \sqrt{F^{\gamma}}$, followed by a second channel operation that is conditioned on the classical register $C$ and uses Kraus operators $\tilde{K}_j^{\gamma} = K_j^{\gamma} (F^{\gamma})^{-1/2}$. To see this, we need only to verify two things: (a) the concatenation of both operations gives the general form and (b) the Kraus operators $\tilde{K}_j^{\gamma}$ for each value of $\gamma$ define a valid CPTP map. The proof of (a) is trivial, and for (b) we need only to verify that $\sum_{j \in \mathcal{I}_{\gamma}} \left(\tilde{K}_j^{\gamma} \right)^{\dagger} \tilde{K}_j^{\gamma} = \mathds{1}_{\gamma}$, where $\mathds{1}_{\gamma}$ is the projector onto the support of $F^{\gamma}$ and $(F^{\gamma})^{-1/2}$ is the corresponding pseudoinverse of $\sqrt{F^{\gamma}}$. We insert the definition to find
\begin{equation}
\begin{aligned}
 \sum_{j \in \mathcal{I}(\gamma)} &\left(\tilde{K}_j^{\gamma} \right)^{\dagger} \tilde{K}_j^{\gamma}\\ 
=& \sum_{j \in \mathcal{I}(\gamma)}(F^{\gamma})^{-1/2} \left(K_j^{\gamma} \right)^{\dagger} K_j^{\gamma} (F^{\gamma})^{-1/2}\\
=& (F^{\gamma})^{-1/2} \left(  \sum_{j \in \mathcal{I}(\gamma)} \left(K_j^{\gamma} \right)^{\dagger} K_j^{\gamma} \right) (F^{\gamma})^{-1/2}\\
=&  (F^{\gamma})^{-1/2} F^{\gamma} (F^{\gamma})^{-1/2} \\
=&  \mathds{1}_{\gamma}.
\end{aligned}
\end{equation}
Clearly, since the general case can thus be considered as a two-step procedure, where the first step gives rise to the announcement $\gamma$ and the second step acts only on Eve's conditional states, it can only strengthen Eve's position by not forcing her to do this second step. Without loss of generality, we can thus assume that Eve's optimal strategy performs only the first step.

Since we assume the sources are protected, Eve cannot have the access to the registers $A$ and $B$ and cannot modify the states in those registers. Therefore, when Eve directly acts on the state $\ket{\Psi}_{ABA'B'}$ shown in Eq. (\ref{eq:sourcereplacement}) from the source-replacement scheme, the joint state $\rho_{ABEC}$ shared by Alice, Bob and Eve along with the classical register $C$ for announcements is as follows:  
\begin{equation}\label{eq:jointstate}
\begin{aligned}
\rho_{ABEC} &= (\mathds{1}_{AB} \otimes \mathcal{E}_{A'B' \rightarrow EC}) (\dyad{\Psi}{\Psi}_{ABA'B'}) \\
&= \sum_{x,y,x',y'} \sqrt{p_x p_{x'} q_y q_{y'}} \dyad{x,y}{x',y'}_{AB} \\ & \otimes \sum_{\gamma}  (\sqrt{F^{\gamma}} \dyad{\varphi_x, \varphi_y}{\varphi_{x'}, \varphi_{y'}} \sqrt{F^{\gamma}})_E\otimes \dyad{\gamma}{\gamma}_C.
\end{aligned}
\end{equation}

\subsection{Key rate evaluation with Devetak-Winter formula}
To distill keys from $\rho_{ABEC}$, Alice and Bob perform measurements using POVMs $M_A$ on the register $A$ and $M_B$ on the register $B$, respectively. Upon measurements, Alice stores her measurement outcomes in a classical register $X$ and Bob stores his in a classical register $Y$. Alice then applies a key map that maps her measurement result in the register $X$ to a raw key bit in the register $K$. We want to point out that the key map step is necessary, but the key map can be trivial, as it is in this PM-MDI QKD protocol. The key map here is an identity map from the register $X$ to the register $K$. Let $\mathcal{G}$ denote the effective CPTP map that transforms $\rho_{ABEC}$ to $\rho_{KYEC}$. In the end, we generate keys from the state $\rho_{KYEC}$, which has the form
\begin{equation}\label{eq:rhoKYEC}
\begin{aligned}
\rho_{KYEC} =& \mathcal{G}(\rho_{ABEC}) \\
=& \sum_{k,y, \gamma}  p(\gamma) p(k,y|\gamma) \dyad{k}{k}_{K}  \otimes \dyad{y}{y}_{Y}  \\ & \otimes  \rho_E^{k,y,\gamma} \otimes \dyad{\gamma}{\gamma}_C,
\end{aligned}
\end{equation}
where $\rho_E^{k, y, \gamma}$ is Eve's conditional state conditioned on Alice holding $k$ in the register $K$, Bob having $y$ in the register $Y$ and the central node announcing $\gamma$. Here, $p(\gamma)$ is a marginal probability of the joint probability distribution $p(k, y, \gamma)$ and $p(k, y |\gamma) = \frac{p(k,y, \gamma)}{p(\gamma)}$ is a conditional probability. 

Under collective attacks, we can evaluate the secret key generation rate using Devetak-Winter formula \cite{Devetak2005}, which is expressed in terms of a single-copy state $\rho_{KYEC}$ shared by Alice, Bob and Eve.

As is typical in the MDI protocols, we can choose to generate keys from each announcement outcome $\gamma$ independently as the announcement is available to all parties. We rewrite $\rho_{ABEC}$ by defining conditional states of Alice, Bob and Eve conditioned on the announcement outcome $\gamma$ as 
\begin{equation}\label{eq:rhoKYEconditionedGamma}
\rho_{KYE}^{\gamma} = \sum_{k,y}  p(k,y|\gamma) \dyad{k}{k}_{K}  \otimes \dyad{y}{y}_{Y}  \otimes  \rho_E^{k,y,\gamma},
\end{equation}
and  $\rho_{KYEC} = \sum_{\gamma} p(\gamma) \rho_{KYE}^{\gamma} \otimes \dyad{\gamma}{\gamma}_C.$

We adapt the Devetak-Winter formula to a general case where the error correction is not necessarily performed at the Shannon limit. In that case, the number of secret bits that we can distill from the state $\rho_{KYE}^{\gamma}$ is $r(\rho_{KYE}^{\gamma})$, which is defined as
\begin{equation}\label{eq:distillablekey_gamma}
r(\rho_{KYE}^{\gamma}) = \max \ [1-\delta_{\text{EC}}^{\gamma} - \chi (K:E)_{\rho_{KYE}^{\gamma}},0],
\end{equation} 
where $\delta_{\text{EC}}^{\gamma}$ is the amount of information leakage per signal during the error correction step for the rounds corresponding to the announcement outcome $\gamma$, and \begin{equation}\label{eq:HolevoInformation}
\chi (K:E)_{\rho_{KYE}^{\gamma}} =S(\rho_E^{\gamma}) - \sum_k p(k|\gamma)S(\rho_E^{k,\gamma})
 \end{equation}
 is the Holevo information, where $S(\rho) = -\Tr(\rho \log_2 \rho)$ is the von Neumann entropy. The states $\rho_{E}^{\gamma}$ and $\rho_{E}^{k,\gamma}$ are defined as:
\begin{equation}\label{eq:defconditionalStates}
\begin{aligned}
 \rho_E^{k,\gamma} &:=\sum_{y} \frac{p(k, y|\gamma)}{p(k|\gamma)} \rho_{E}^{k, y, \gamma} \\
 & = \sum_{y} p(y|k,\gamma) \rho_{E}^{k, y, \gamma}, \\
 \rho_{E}^{\gamma} &:= \sum_{k}p(k|\gamma)\rho_E^{k,\gamma}.
 \end{aligned}
 \end{equation}

In the Shannon limit, we have $1-\delta_{\text{EC}}^{\gamma} = I(K:Y)_{\rho_{KYE}^{\gamma}}$, where $I(K:Y)_{\rho_{KYE}^{\gamma}}$ is the classical mutual information, and thus we recover the original Devetak-Winter formula in Eq. (\ref{eq:distillablekey_gamma}). Another important observation is that $\delta_{\text{EC}}^{\gamma}$ is directly determined from the experimentally observed correlations. 

The total number of secret bits that we can distill from the state $\rho_{KYEC}$, denoted by $\tilde{r}(\rho_{KYEC})$, is defined as
\begin{equation}\label{eq:totalkeyrate}
\tilde{r}(\rho_{KYEC}) = \sum_{\gamma} p(\gamma) r(\rho_{KYE}^{\gamma}).
\end{equation}

From Eq. (\ref{eq:jointstate}), we can calculate Eve's conditional states $\rho_E^{k, y, \gamma}$ as
\begin{equation}\label{eq:rhoEkygamma}
\rho_E^{k, y, \gamma} =\dyad{\Theta^{\gamma}_{k,y}}{\Theta^{\gamma}_{k,y}},
\end{equation}
where we define 
\begin{equation}\label{eq:Thetakygamma}
\ket{\Theta^{\gamma}_{k,y}} =\frac{\sqrt{F^{\gamma}} \ket{\varphi_k, \varphi_y}}{\sqrt{\bra{\varphi_k, \varphi_y}F^{\gamma}\ket{\varphi_k, \varphi_y}}}.
\end{equation}
Then, by substituting Eq. (\ref{eq:rhoEkygamma}) into Eq. (\ref{eq:defconditionalStates}), we can calculate the conditional states $\rho_{E}^{\gamma}$ and $\rho_{E}^{k,\gamma}$, and evaluate $\chi (K:E)_{\rho_{KYE}^{\gamma}}$ in Eq. (\ref{eq:HolevoInformation}) to obtain $r(\rho_{KYE}^{\gamma})$ in Eq. (\ref{eq:distillablekey_gamma}).

From the relation between $\rho_{ABEC}$ and $\{F^{\gamma}\}$ shown in Eq. (\ref{eq:jointstate}), we notice that a full knowledge of $\{F^{\gamma}\}$ gives us a full knowledge of $\rho_{ABEC}$ and thus we can determine the key rate using Eq. (\ref{eq:totalkeyrate}). However, if we cannot uniquely determine $F^{\gamma}$, then we cannot uniquely determine $\rho_{ABEC}$. In that case, we have a set of compatible density operators $\rho_{ABEC}$, that is, $\mathscr{C}=\{\rho_{ABEC}: \rho_{ABEC} \text{ is compatible with experimental observations}\}$. Thus, we need to consider the worst-case scenario by taking the minimum of $\tilde{r}$ over the set $\mathscr{C}$, or equivalently, over the set $\mathcal{C}$ =  $\{\rho_{KYEC}: \rho_{KYEC} = \mathcal{G}(\rho_{ABEC}), \text{ where }\rho_{ABEC} \in \mathscr{C}\}$. 
 
In this situation, the asymptotic key rate $R^{\infty}$ should be expressed as 
\begin{equation}\label{eq:keyrate_minization}
\begin{aligned}
R^{\infty} &= \min_{\rho_{KYEC} \in \mathcal{C}} \tilde{r}(\rho_{KYEC})\\
&=\min_{\rho_{ABEC} \in \mathscr{C}} \tilde{r}(\mathcal{G}(\rho_{ABEC})).
\end{aligned}
\end{equation}

The essential part of the optimization is to optimize the Holevo information $\chi(K:E)$ by finding the all possible Eve's conditional states, which are needed to evaluate Eq. (\ref{eq:HolevoInformation}).

We remark that most of the discussion so far is general to a generic MDI QKD protocol. In the next section, we will adapt this procedure to our specific PM-MDI QKD protocol.

\subsection{Determination of Eve's POVM for PM-MDI QKD}\label{sec:POVMDetermination}

As discussed in the previous sections, knowing Eve's POVM elements allows us to calculate the key rate, since the minimization in Eq. (\ref{eq:keyrate_minization}) is now over a set containing only one element. We will now explain how our choice of test states (coherent states with a continuum of complex amplitudes) allows in principle to determine Eve's POVM elements. 

For simplicity, let us concentrate on the case of testing a measurement device acting on a single mode (rather than the two-mode case of our protocol). Knowing some POVM element $\tilde{F}$ is equivalent to being able to predict the probability $p(\tilde{F})$ of the associated outcome for any input state $\rho$ as $p(\tilde{F}) = \Tr[ \rho \; \tilde{F}]$. We can now use the phase-space formalism of quantum mechanics (see, for example, Refs. \cite{Cahill1969a,Cahill1969b})  where we use the $P$-function  representation of $\rho = \int d^2 \alpha\;  P(\alpha) \dyad{\alpha}{\alpha}$ so that we have
\begin{equation}
p(\tilde{F}) = \int d^2 \alpha \; P(\alpha) \; \bra{\alpha} \tilde{F} \ket{ \alpha}  \; .
\end{equation}
As we see from this equation, knowledge of the function $p( \tilde{F} |\alpha) :=\bra{\alpha}  \tilde{F}  \ket{\alpha} $ allows the prediction of $p( \tilde{F} )$ for all input states for which the $P$ function of the density matrix $\rho$ exists. So testing the measurement device with all possible coherent states $\ket{\alpha}$ and observing the corresponding probabilities $p( \tilde{F} |\alpha)$ is equivalent to knowing $ \tilde{F} $. 

Actually, using results from \cite{Cahill1969a,Cahill1969b}, one can reconstruct the operator $ \tilde{F} $ explicitly also in cases where the $P$ function of $\rho$ may not exist. Let us go through the arguments directly for the POVM elements $F^{\gamma}$ for the outcome $\gamma$ in the two-mode case. We adapt the Eqs (3.4)-(3.6) from Ref. \cite{Cahill1969b} to our scenario. 

By substituting Eqs. (3.4) and (3.6) into Eq. (3.5) from Ref. \cite{Cahill1969b}, we obtain a power series for each $F^{\gamma}$ as:

\small
\begin{equation}
\begin{aligned}
F^{\gamma} =\sum_{\substack{n_1,n_2, \\ m_1,m_2=0}}^{\infty} &\frac{ \Big(\partial_{\alpha_1}^{m_1}\partial_{\alpha_2}^{m_2}\partial_{\bar{\alpha}_1}^{n_1}\partial_{\bar{\alpha}_2}^{n_2} \bra{\alpha_1,\alpha_2}F^{\gamma} \ket{\alpha_1,\alpha_2}\Big)|_{\alpha_1=0,\alpha_2=0}}{m_1! m_2! n_1! n_2!} \\
&\times  ({a_1}^{\dagger})^{n_1} ({a_2}^{\dagger})^{n_2} {a_1}^{m_1} {a_2}^{m_2},
\end{aligned}
\end{equation}
where  $\alpha_1, \alpha_2$ and their complex conjugated counterparts $\bar{\alpha}_1,\bar{\alpha}_2$ are treated as independent variables, and $a_1, {a_1}^\dagger$, $a_2, {a_2}^\dagger$ are the annihilation and creation operators of the two modes. Since $F^{\gamma}$ is a POVM element and thus has bounded eigenvalues, such series exist and converge \cite{Cahill1969a}. Using the two-mode test states $|\alpha_1,\alpha_2\rangle$ and the associated observed probabilities $p(\gamma|\alpha_1,\alpha_2) =\bra{\alpha_1,\alpha_2}F^{\gamma} \ket{\alpha_1,\alpha_2}$ thus uniquely determines $F^\gamma$.  

\normalsize

Note that a full description of $F^{\gamma}$ as shown above is more than what we actually need since we are only interested in how $F^{\gamma}$ acts on the subspace $\spn(\mathcal{S})$, which is only a four-dimensional space.

For this, we need to be able to calculate off-diagonal elements of the form $\bra{\alpha_1,\alpha_2}F^{\gamma} \ket{\beta_1,\beta_2}$. It is an interesting question of whether we can estimate these elements well enough with just a few number of coherent states. (The diagonal elements are directly accessible.) We present now the handle to attack this question.

We first notice that characterizing $F^\gamma$ on $\spn(\mathcal{S})$ is equivalent to the question whether the operator $\dyad{\beta_1, \beta_2}{\alpha_1,\alpha_2}$ can be approximated to arbitrary precision in the Hilbert-Schmidt norm by the discrete diagonal coherent state representation \cite{Mukunda1978, Sharma1981}:
\begin{equation}
\dyad{\beta_1, \beta_2}{\alpha_1,\alpha_2}  = \sum_{i=1}^{\infty} \lambda_{i} \dyad{\omega_1^{(i)},\omega_2^{(i)}}{\omega_1^{(i)}, \omega_2^{(i)}},
\end{equation} 
where we use sets of tensor products of coherent states $ \ket{\omega_1^{(i)},\omega_2^{(i)}}$  and complex numbers $\lambda_i \in \mathbb{C}$. 

Then, we can write $\bra{\alpha_1,\alpha_2}F^{\gamma} \ket{\beta_1,\beta_2}$ as a sum of observed values $\bra{\omega_1^{(i)},\omega_2^{(i)}}F^{\gamma}\ket{\omega_1^{(i)},\omega_2^{(i)}}$ as
\begin{equation}
\begin{aligned}
\bra{\alpha_1,\alpha_2}F^{\gamma} \ket{\beta_1,\beta_2} &= \Tr(F^{\gamma}\dyad{\beta_1, \beta_2}{\alpha_1,\alpha_2})\\
&=\sum_{i=1}^{\infty} \lambda_i \bra{\omega_1^{(i)},\omega_2^{(i)}}F^{\gamma}\ket{\omega_1^{(i)},\omega_2^{(i)}}.
\end{aligned}
\end{equation}

By appropriate choices of $\{\ket{\omega_1^{(i)},\omega_2^{(i)}}\}_{i=1}^{N}$, we will be able to get a good approximation by terminating the summation at $N$. From the approximation, we will then determine a set of POVMs compatible with experimental correlations, which is a neighborhood of the POVM that Eve actually performed. When we calculate the key rate in this case, we need to perform the minimization in Eq. (\ref{eq:keyrate_minization}). In that case, we may apply numerical methods \cite{Coles2016, Winick2018} to perform the desired optimization. If such an approximation makes this set of compatible POVMs small enough, then the key rate with several choices of test states would be close to the key rate with infinite choices of test states. We leave the detailed analysis of finite choices of test states scenario to the future work.

In Appendix \ref{sec:representation}, we will discuss how to represent $F^{\gamma}$ in the four-dimensional subspace $\spn(\mathcal{S})$ after knowing $\bra{\alpha_1,\alpha_2}F^{\gamma} \ket{\beta_1,\beta_2}$ for $\ket{\alpha_1,\alpha_2}, \ket{\beta_1,\beta_2} \in \mathcal{S}$.

\section{SIMULATION}\label{sec:simulation}
We perform simulations to study the loss scaling of this PM-MDI QKD protocol and also the stability of the protocol.

\subsection{Loss-only scenario}
To show that the key rate of this protocol has a scaling of $\sqrt{\eta}$ with the single-photon transmissivity $\eta$ between Alice and Bob, we first study the loss-only scenario. We simulate the quantum channel as a lossy channel and we consider the normal situation where Charlie (Eve) performs the measurements so that the observed statistics during the parameter estimation step is compatible with Charlie performing the measurement shown in Fig. \subref*{fig:scheme_a}. That is, we calculate the POVM $F$ corresponding to the real setup. Our protocol can verify via test states in the test mode that this is the actual POVM performed by Eve in the loss-only scenario. For the purpose of our presentation, we consider a symmetric setup, that is, Charlie is at an equal distance from Alice and Bob, and the loss in each path is the same. For a total transmissivity $\eta$ between Alice and Bob, each segment has the transmissivity $\sqrt{\eta}$.

In this situation, when Alice sends a coherent state $\ket{\alpha_A}$ and Bob sends a coherent state $\ket{\alpha_B}$ in the same optical mode, the state becomes $\ket{\sqrt{\sqrt{\eta}}\alpha_A, \sqrt{\sqrt{\eta}}\alpha_B}$ after the lossy channel. When Charlie performs the measurement on this state, the probability for each announcement outcome $\gamma$ can be calculated as follows:

\small
\begin{equation}
\begin{aligned}
\bra{\alpha_A, \alpha_B} F^+ \ket{\alpha_A, \alpha_B} &= (1-e^{-\frac{\sqrt{\eta}\abs{\alpha_A + \alpha_B}^2}{2}})e^{-\frac{\sqrt{\eta}\abs{\alpha_A - \alpha_B}^2}{2}}, \\
\bra{\alpha_A, \alpha_B} F^- \ket{\alpha_A, \alpha_B} &=  e^{-\frac{\sqrt{\eta}\abs{\alpha_A + \alpha_B}^2}{2}}(1-e^{-\frac{\sqrt{\eta}\abs{\alpha_A - \alpha_B}^2}{2}}),\\
\bra{\alpha_A, \alpha_B} F^? \ket{\alpha_A, \alpha_B} &= e^{-\frac{\sqrt{\eta}\abs{\alpha_A + \alpha_B}^2}{2}}e^{-\frac{\sqrt{\eta}\abs{\alpha_A - \alpha_B}^2}{2}}, \\
\bra{\alpha_A, \alpha_B} F^d \ket{\alpha_A, \alpha_B} &= (1-e^{-\frac{\sqrt{\eta}\abs{\alpha_A + \alpha_B}^2}{2}})(1-e^{-\frac{\sqrt{\eta}\abs{\alpha_A - \alpha_B}^2}{2}}).
\end{aligned}
\end{equation}
\normalsize

Specifically, the conditional probability of each announcement outcome for each state in the set $\mathcal{S}$ is summarized in Table \ref{table:probabilitydistribution}. From this table, we can directly evaluate the classical mutual information $I(K:Y)$ as
\begin{equation}
\begin{aligned}
I(K:Y)_{\rho_{KYE}^{+}} & = I(K:Y)_{\rho_{KYE}^{-}} &=1,\\
I(K:Y)_{\rho_{KYE}^{?}} & = I(K:Y)_{\rho_{KYE}^{d}}  &=0.
\end{aligned}
\end{equation}

Clearly, we cannot distill keys from $\gamma = ``?"$ and $\gamma = ``d"$ announcements. Also, we find $\delta_{\text{EC}}^{+} = \delta_{\text{EC}}^{-}=0$ since no error correction is needed in this loss-only scenario. Now, we only need to evaluate $\chi(K:E)$ for $\gamma = ``+"$ and $\gamma = ``-"$. We first find conditional states $\rho_{E}^{k,y,+}$ and $\rho_{E}^{k,y,-}$ defined in Eq. (\ref{eq:rhoEkygamma}).

\begin{table}\small 
\begin{center}
\caption{\label{table:probabilitydistribution} Conditional probability distribution of announcement outcomes given the states from $\mathcal{S}$ in the loss-only scenario. $\eta$ is the single-photon transmissivity between Alice and Bob and $\mu$ is the intensity of coherent states in the key-generation mode.}
\begin{tabular}{ c|c|c|c|c }
 \hline
 \hline
$\alpha_A, \alpha_B$& $ +\sqrt{\mu},+ \sqrt{\mu}$ & $ -\sqrt{\mu},  -\sqrt{\mu}$ & $+\sqrt{\mu}, -\sqrt{\mu}$ & $ -\sqrt{\mu}, + \sqrt{\mu}$ \\
\hline
$p(+|\alpha_A,\alpha_B)$ & $1-e^{-2\sqrt{\eta}\mu}$& $1-e^{-2\sqrt{\eta}\mu}$ & 0 &0 \\

 $p(-|\alpha_A,\alpha_B)$ & 0& 0& $1-e^{-2\sqrt{\eta}\mu}$ &$1-e^{-2\sqrt{\eta}\mu}$ \\

 $p(?|\alpha_A,\alpha_B)$ & $e^{-2\sqrt{\eta}\mu}$ & $e^{-2\sqrt{\eta}\mu}$& $e^{-2\sqrt{\eta}\mu}$ &$e^{-2\sqrt{\eta}\mu}$ \\

 $p(d|\alpha_A,\alpha_B)$ & 0& 0& 0 &0 \\
 \hline
 \hline
\end{tabular}
\end{center}
\end{table}
As we can notice from Table \ref{table:probabilitydistribution}, in the loss-only scenario, whenever Alice and Bob prepare coherent states with a $\pi$ phase difference, Charlie will never announce $\gamma=``+"$ and whenever they prepare coherent states with the same phase, Charlie will never announce $\gamma=``-"$. Because $p(0,1,+)=p(1,0,+) =0$ and $p(0,0,-)=p(1,1,-) =0$, each of the states $\rho_E^{k, +}$ and $\rho_E^{k, -} \ \forall k \in \{0,1\}$ is a pure state so that $S(\rho_E^{k, +}) = S(\rho_E^{k, -}) =0$. Therefore, we only need to evaluate $S(\rho_E^{+})$ and $S(\rho_E^{-})$. 


In this loss-only case, 
\begin{equation}
\begin{aligned}
\rho_{E}^+ &= \frac{1}{2}(\dyad{\Theta_{0,0}^{+}} + \dyad{\Theta_{1,1}^{+}}), \\
\rho_{E}^- &= \frac{1}{2}(\dyad{\Theta_{0,1}^{-}} + \dyad{\Theta_{1,0}^{-}}).
\end{aligned}
\end{equation}

The eigenvalues of $\rho_{E}^+$ are $\frac{1}{2}(1\pm \abs{\bra{\Theta_{0,0}^{+}}\ket{\Theta_{1,1}^+}})$ and thus $S(\rho_E^{+}) = h(\frac{1 - \abs{\bra{\Theta_{0,0}^{+}}\ket{\Theta_{1,1}^+}}}{2})$, where $h(x) = -x \log_2(x) - (1-x) \log_2(1-x)$ is the binary entropy function. 
Similarly, the eigenvalues of $\rho_{E}^-$ are $\frac{1}{2}(1\pm \abs{\bra{\Theta_{0,1}^{-}}\ket{\Theta_{1,0}^-}})$ and thus $S(\rho_E^{-}) = h(\frac{1 - \abs{\bra{\Theta_{0,1}^{-}}\ket{\Theta_{1,0}^-}}}{2})$. 
Using the definition of $\ket{\Theta_{k,y}^{\gamma}}$ in Eq. (\ref{eq:Thetakygamma}), we obtain 

\begin{widetext}
 \begin{equation}
 \begin{aligned}
\braket{\Theta_{0,0}^{+}}{\Theta_{1,1}^+} &= \frac{\bra{+\sqrt{\mu},+\sqrt{\mu}}F^+\ket{-\sqrt{\mu},-\sqrt{\mu}}}{\sqrt{\bra{+\sqrt{\mu},+\sqrt{\mu}}F^+\ket{+\sqrt{\mu},+\sqrt{\mu}}\bra{-\sqrt{\mu},-\sqrt{\mu}}F^+\ket{-\sqrt{\mu},-\sqrt{\mu}}}}, \\
\braket{\Theta_{0,1}^{-}}{\Theta_{1,0}^-} &= \frac{\bra{+\sqrt{\mu},-\sqrt{\mu}}F^-\ket{-\sqrt{\mu},+\sqrt{\mu}}}{\sqrt{\bra{+\sqrt{\mu},-\sqrt{\mu}}F^-\ket{+\sqrt{\mu},-\sqrt{\mu}}\bra{-\sqrt{\mu},+\sqrt{\mu}}F^-\ket{-\sqrt{\mu},+\sqrt{\mu}}}}.
\end{aligned}
\end{equation}
\end{widetext}
Thus, we have $S(\rho_E^{+}) = S(\rho_E^{-}) = h(\frac{1-e^{-4 \mu (1-\sqrt{\eta})}e^{-2 \mu \sqrt{\eta}}}{2})$. We provide explicit expressions of $F^{\gamma}$ for this loss-only scenario in Sec. \ref{sec:simulationdetails:loss} of Appendix \ref{sec:simulationdetails}, using which the reader can check the result directly.

Finally, we obtain the expression of secret key generation rate as a function of $\eta$ and the intensity $\mu$ in this loss-only scenario as
 \begin{equation}\label{eq:losskeyrate}
R^{\infty} = (1-e^{-2 \mu \sqrt{\eta} }) \Big[1- h\Big(\frac{1-e^{-4 \mu (1-\sqrt{\eta})}e^{-2 \mu \sqrt{\eta}}}{2}\Big)\Big].
\end{equation}

For small values of $\eta$, $R^{\infty} \approx 2\mu (1-h(\frac{1-e^{-4\mu}}{2})) \sqrt{\eta}$. When we take the optimal value of $\mu$, which is $\mu_{\text{opt}} \approx 0.1146$, then we find $R^{\infty} \approx 0.0714 \sqrt{\eta}$, thus confirming the rate scaling of $R^{\infty} = O(\sqrt{\eta})$.

In Fig. \ref{fig:keyrate}, the short-dashed line is the asymptotic key rate of this loss-only scenario as a function of the transmission distance $L$, where we take $\eta = 10^{-\frac{0.2L}{10}}$ and $\mu$ is optimized for each distance $L$. The solid line is the fundamental repeaterless bound $-\log_2(1-10^{-\frac{0.2 L}{10}})$ \cite{Pirandola2017}. This calculation gives an intuitive understanding on how the PM-MDI QKD  can beat the repeaterless key rate bound. We see that this PM-MDI QKD protocol beats the repeaterless bound at around $150$ km. Our key rate expression in Eq. (\ref{eq:losskeyrate}) is tight for the loss-only scenario. Therefore, we expect this is the loss limit for PM-MDI QKD.

\subsection{Realistic Imperfections}
It is of practical interest to study how stable this protocol is in noisy scenarios. In particular, we simulate the scenario with realistic imperfections in experimental devices, including the dark counts of detectors, mode mismatch and phase mismatch, detector inefficiency, and error correction inefficiency. In this section, we briefly introduce sources of imperfections and corresponding simulation parameters, then explain the correlations that Alice and Bob would observe in our simulation model, and finally present the results of our key rate calculation. In Appendix \ref{sec:simulationdetails}, we provide more detailed explanations for the physical model of each imperfection. 

For the purpose of presentation, we assume that both detectors have the same detector efficiency $\eta_d$ and the same dark count probability $p_d$. We remark that the simulation method described in Appendix \ref{sec:simulationdetails} is also applicable to more general situations.

In the ideal implementation of this protocol, Alice and Bob should prepare coherent states in the same optical mode, that is, with the same spectral, temporal profiles and the same polarization, in order to have single-photon interference at the beam splitter. In reality, since their states may come from different lasers and pass through different optical components before reaching the central node, the modes of their states are not necessarily perfectly matched. Thus, we consider the relative mode mismatch between their states with a simulation parameter $V$. In our simulation, if without any mode mismatch, the state arriving at the central node from Alice and Bob is supposed to be $\ket{\alpha_A,\alpha_B}$, then with the mode mismatch, the state becomes $\ket{\alpha_A,\sqrt{V}\alpha_B}$ in the original mode and $\ket{0,\sqrt{1-V}\alpha_B}$ in a second mode. Both modes enter Charlie's devices independently.

Another source of imperfection considered in our simulation model is the phase mismatch. 
In the key-generation mode, Alice and Bob are supposed to prepare states in the set $\mathcal{S}$, which are coherent states with the same global phase and with the encoding information in the relative phases. In reality, the global phase is not guaranteed to be the same when states reach the detectors. Therefore, we consider the situation where there is a relative phase mismatch between Alice's signal state and Bob's signal state. If without any phase mismatch, the state is supposed to be $\ket{\alpha_A,\alpha_B}$, then due to the phase mismatch, the state is changed to $\ket{\alpha_A,\alpha_B e^{i \delta}}$ with a simulation parameter $\delta$.

\begin{figure}
\includegraphics[width=0.5\textwidth]{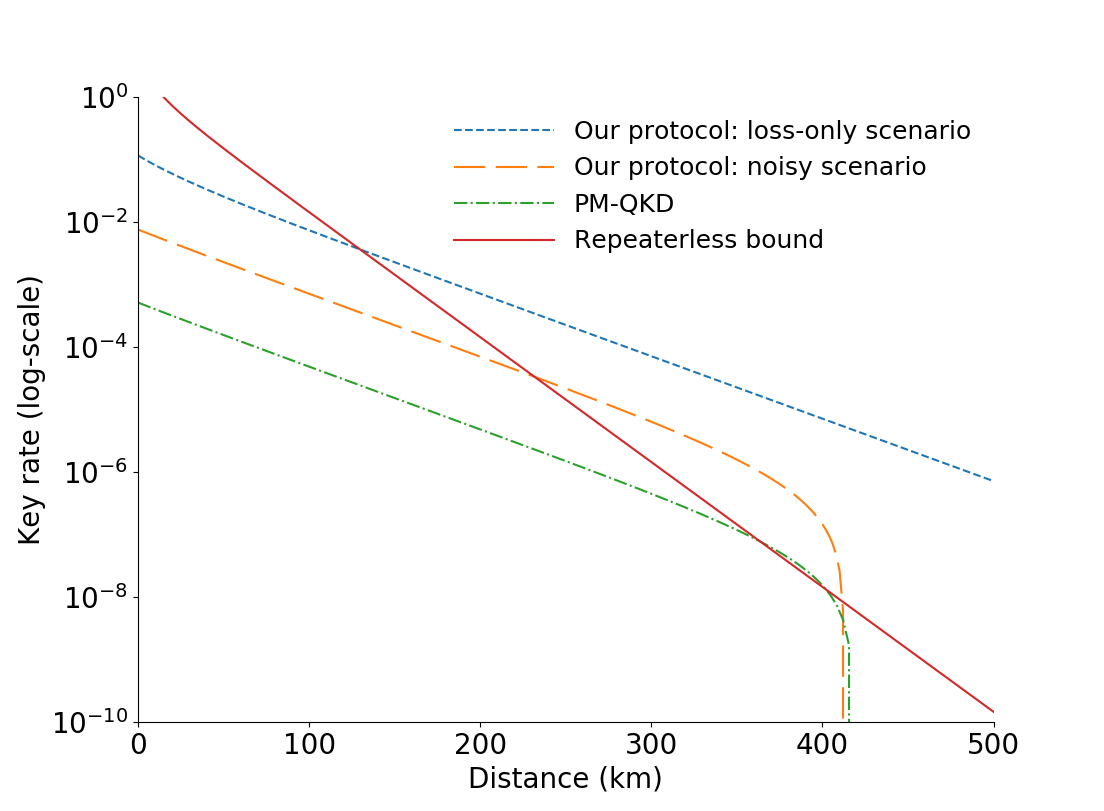}
 \caption{\label{fig:keyrate} A log-linear plot for the key rate as a function of transmission distance. The solid line is the fundamental repeaterless secret key capacity bound $-\log_2(1-10^{-\frac{0.2 L}{10}})$ for a transmission distance $L$ (in km) \cite{Pirandola2017}. The short-dashed line is the key rate for PM-MDI QKD in the loss-only scenario in Eq. (\ref{eq:losskeyrate}). The long-dashed line is the key rate for PM-MDI QKD with experimentally feasible parameters listed in Table \ref{table:simulationParamters}. The dash-dotted line is phase-matching QKD key rate provided by authors of \cite{Ma2018} with the same set of experimental parameters. The intensity $\mu$ is optimized for each distance in these curves.}
 \end{figure}

Table \ref{table:simulationParamters} lists the choice of parameters in our simulation. In particular, we choose the same values for the efficiency of a detector $\eta_d$ and the dark count probability of a detector $p_d$ as those used in Ref. \cite{Ma2018} for the comparison purpose. We also select pessimistic values for the mode mismatch and phase mismatch to demonstrate the feasibility of beating the repeaterless bound with currently available devices.

\begin{table}
\begin{center}
\caption{\label{table:simulationParamters}Values for simulation parameters. They are experimentally feasible and might be pessimistic values. See main text for more explanations.}

\begin{tabular*}{0.48\textwidth}{@{\extracolsep{\fill}}lcc } 
 \hline
 \hline
 Detector efficiency $\eta_d$ & &  14.5\% \\ 

  Detector dark count probability $p_d$& & $8 \times 10^{-8}$  \\ 

 Mode mismatch ($1-V$)& &  5\%  \\ 

Phase mismatch $\delta$ & &   $\frac{\pi}{60}$  \\ 

Error correction efficiency $f_{\text{EC}}$ & &  1.15  \\ 
 \hline
 \hline
\end{tabular*}
\end{center}
\end{table}

We now give the expressions for the probability of each announcement outcome $\gamma$ given each choice of the input state in terms of the simulation parameters $V$, $\delta$, $\eta_d$, and $p_d$. We define the total transmissivity as $\eta = \eta_{t}\eta_{d}^2$, where $\eta_t$ is the channel transmission probability between Alice and Bob.

\small
\begin{equation}\label{eq:simulationcorrelation}
\begin{aligned}
\bra{\alpha_A, \alpha_B} F^+ \ket{\alpha_A, \alpha_B} &=(1-p_d) (1-\xi_1\xi_2)\xi_2\xi_3 \\
 &+(1-p_d)p_d \xi_1\xi_2^2 \xi_3,\\
\bra{\alpha_A, \alpha_B} F^- \ket{\alpha_A, \alpha_B} &=  (1-p_d)\xi_1\xi_2 (1-\xi_2\xi_3) \\
&+ (1-p_d)p_d \xi_1\xi_2^2 \xi_3, \\
\bra{\alpha_A, \alpha_B} F^? \ket{\alpha_A, \alpha_B} &= (1-p_d)^2\xi_1\xi_2^2\xi_3,\\
\bra{\alpha_A, \alpha_B} F^d \ket{\alpha_A, \alpha_B} &= p_d (1-\xi_1\xi_2)\xi_2\xi_3+p_d \xi_1\xi_2 (1-\xi_2\xi_3)\\
&+p_d^2\xi_1\xi_2^2\xi_3+(1-\xi_1\xi_2)(1-\xi_2\xi_3),
\end{aligned}
\end{equation}
\normalsize
where for the simplicity of writing, we have made the following definitions
\begin{equation}
\begin{aligned}
\xi_{1}&=e^{-\frac{1}{2}\sqrt{\eta}\abs{\alpha_A + \sqrt{V}\alpha_B e^{i\delta}}^2},\\
\xi_{2}&=e^{-\frac{1}{2}\sqrt{\eta}(1-V)\abs{\alpha_B}^2}, \\
\xi_{3} &=e^{-\frac{1}{2}\sqrt{\eta}\abs{\alpha_A - \sqrt{V}\alpha_B e^{i\delta}}^2}.
\end{aligned}
\end{equation}

From Eq. (\ref{eq:simulationcorrelation}), it is straightforward to derive the conditional probability of each announcement outcome given the state in $\mathcal{S}$. Similar to the loss-only scenario, we also discover that the mutual information $I(K:Y)$ is zero for $\gamma =`` \ ? \ "$ and $\gamma = `` \ d \ "$ since the probability of making those announcements is independent from the signal states sent by Alice and Bob in our simulation. Thus, we only generate keys from $\gamma = ``+"$ and $\gamma = ``-"$ outcomes. 

We define error rates $e_{+}$ and $e_{-}$ given the announcement outcome $\gamma = ``+"$ and $\gamma = ``-"$, respectively. 
\begin{equation}
\begin{aligned}
e_{+} &:= p(0,1|+)+p(1,0|+) \\
&=\frac{ \zeta_2 - (1-p_d)\zeta_1\zeta_2}{\zeta_1 + \zeta_2 - 2(1-p_d)\zeta_1\zeta_2}, \\
e_{-} &:= p(0,0|-)+p(1,1|-) \; = e_{+},
\end{aligned}
\end{equation}
where we define $\zeta_1 = e^{-\sqrt{\eta} \mu (1-\sqrt{V}\cos(\delta))}$ and $\zeta_2 =  e^{-\sqrt{\eta} \mu (1+\sqrt{V}\cos(\delta))}$.

To take the inefficiency of error correction into consideration, we take the following values for $\delta_{\text{EC}}^{+}$ and $\delta_{\text{EC}}^{-}$:
\begin{equation}\label{eq:simulationclassical}
\begin{aligned}
\delta_{\text{EC}}^{+}&=f_{\text{EC}} \ h(e_+),\\
\delta_{\text{EC}}^{-}&= f_{\text{EC}} \ h(e_-) \; =  \delta_{\text{EC}}^{+},
\end{aligned}
\end{equation}
where $f_{\text{EC}}$ is the efficiency of error correction.

The rest of the task is to find each of $\rho_{E}^{k,y,+}$ and $\rho_{E}^{k,y,-}$. In Appendix \ref{sec:simulationdetails}, we give explicit expressions for $F$ in this scenario. Using Eqs. (\ref{eq:simulationPOVM}) and (\ref{eq:statesVector}), we can find the four-dimensional representation of each of $\rho_{E}^{k,y,+}$ and $\rho_{E}^{k,y,-}$. We numerically evaluate the Holevo information $\chi(K:E)$ for $\gamma = ``+"$ and $\gamma = ``-"$ (even though it is still possible but non-trivial to evaluate $\chi(K:E)$ analytically). 

In Fig. \ref{fig:keyrate}, the long-dashed line shows the result of our simulation. With those experimentally feasible parameters, we see this PM-MDI QKD protocol can still beat the repeaterless bound and this crossover happens at around $250$ km. We also compare our results with the key rates from Ref. \cite{Ma2018} for the same set of experimental parameters. We notice a distinct gap that we attribute to the tomographically complete set of test states and to the fact that our proof technique is tight. Both approaches can provide stronger advantages over the repeaterless bound for less pessimistic experimental parameters.

\section{SUMMARY AND OUTLOOK}\label{sec:outlook}
We have presented a variation of PM-MDI QKD protocol, which uses coherent states (without phase randomization) as test states to probe Eve's attacks and uses exactly one phase-matching pair of coherent states as signal states for key generation. From the tomographical completeness of test states, we can uniquely determine Eve's measurement POVM and then derive Eve's conditional states needed for the direct evaluation of the Devetak-Winter formula (no minimization). We calculate the asymptotic key rate of this PM-MDI QKD protocol against collective attacks in the scenario of infinite choices of test states.  Our analytical key rate formula for the loss-only scenario confirms that the PM-MDI QKD protocol has $O(\sqrt{\eta})$ scaling, better than $O(\eta)$ scaling of the fundamental repeaterless bound \cite{Pirandola2017}. It also shows the loss limit of PM-MDI QKD. In addition, we simulate the key rate with experimentally feasible parameters (if not even pessimistic), and show that PM-MDI QKD can beat the repeaterless bound between roughly 250 km and 400 km. An interesting question for the future work is whether we can approach this key rate bound by using a few choices of test states. We also presented a possible path to proceed with such an analysis. Such an analysis needs to include the study of optimal choices of test states and the effects on the key rate due to nonunique determination of Eve's POVM. We also believe the generalization of decoy state idea presented in this paper can be useful for other QKD protocols.

\textit{Note added.} During the preparation of this paper, we have presented our protocol, proof idea and main results in a conference \cite{Lin2018}. After our presentation, two other works considering similar protocols were posted on a preprint server \cite{Cui2018, Curty2018}. In this article, we compared differences among those similar protocols and showed the clear distinction. Our initial idea was conceived independently from these two works and our analysis has already finished before the appearance of these two works. We nevertheless include references to these publications in our paper for the convenience of the readers.

\begin{acknowledgments}
We thank Ashutosh Marwah, Barry Sanders and Kei Nemoto for helpful discussions regarding the tomographic reconstruction of the POVM elements, and Pei Zeng for discussion about the security proof in Ref. \cite{Ma2018} and for providing data for the key rate curve of PM-QKD in Fig. \ref{fig:keyrate}. The work has been performed at the Institute for Quantum Computing, University of Waterloo, which is supported by Industry Canada. The research has been supported by NSERC under the Discovery Program, Grant No. 341495. Financial support for this work has been partially provided by Huawei Technologies Canada Co., Ltd.
\end{acknowledgments}

\appendix

\section{A REPRESENTATION OF EVE'S POVM IN THE SUBSPACE $\mathcal{S}$}\label{sec:representation}
In this Appendix, we will describe how to find a representation of Eve's POVM $\{F^{\gamma}\}$ in an orthonormal basis of the subspace spanned by the signal states, which we previously denoted as $\spn(\mathcal{S})$. When we discuss about the two-mode coherent states $\ket{\alpha_A,\alpha_B}$ prepared by Alice and Bob, for the ease of notation, we write $\vec{\alpha} = (\alpha_A,\alpha_B)$ and $\ket{\vec{\alpha}}=\ket{\alpha_A,\alpha_B}$. 

If we are given $\bra{\vec{\alpha}}F^{\gamma}\ket{\vec{\beta}}$ for every $\vec{\alpha} ,\vec{\beta} \in \mathcal{S}$, then the procedure described here allows us to find a four-dimensional representation of $F^{\gamma}$ in the subspace $\spn(\mathcal{S})$ and helps us evaluate the von Neumann entropy of Eve's conditional states more straightforwardly. We remark here that the values of $\bra{\vec{\alpha}}F^{\gamma}\ket{\vec{\beta}}$ can be determined by test states in the test mode of our protocol, as discussed in Sec. \ref{sec:POVMDetermination}. For our simulations in Appendix \ref{sec:simulationdetails}, we also provide a simulation method to obtain $\bra{\vec{\alpha}}F^{\gamma}\ket{\vec{\beta}}$.

Before we proceed, we want to emphasize that the set $\mathcal{S}$ defined in Eq. (\ref{eq:definitionS}) is a basis for the subspace $\spn(\mathcal{S})$ and we will use this particular ordering of basis elements in the later discussion.

\subsection{Orthonormal basis decomposition}
Since two coherent states $\{\ket{+\sqrt{\mu}} , \ket{-\sqrt{\mu}}\}$ span a two-dimensional space, we start with a canonical two-dimensional description of $\{\ket{+\sqrt{\mu}} , \ket{-\sqrt{\mu}}\}.$
\begin{equation}\label{eq:decomposition}
\begin{aligned}
\ket{+\sqrt{\mu}} &= c_0 \ket{e_0} + c_1 \ket{e_1}, \\
\ket{-\sqrt{\mu}} &= c_0 \ket{e_0} - c_1 \ket{e_1}, 
\end{aligned}
\end{equation}
where $\{\ket{e_0},\ket{e_1}\}$ is an orthonormal basis,  $\abs{c_0}^2 +\abs{c_1}^2 =1$ and $\abs{c_0}^2 -\abs{c_1}^2 = \braket{+\sqrt{\mu}}{-\sqrt{\mu}}$. Without loss of generality, we choose $c_0$ and $c_1$ to be real numbers by absorbing the complex phases into the definitions of $\ket{e_0}, \ket{e_1}$. We remark here that the explicit expressions for $\ket{e_0}$ and $\ket{e_1}$ are irrelevant for our discussion, but a canonical choice of this basis written in the Fock state basis is 
\begin{equation}
\begin{aligned}
\ket{e_0} &= \frac{1}{\sqrt{\cosh(\mu)}}\sum_{n=0}^{\infty}\frac{\sqrt{\mu}^{2n}}{\sqrt{(2n)!}}\ket{2n},\\
\ket{e_1} &= \frac{1}{\sqrt{\sinh(\mu)}}\sum_{n=0}^{\infty}\frac{\sqrt{\mu}^{2n+1}}{\sqrt{(2n+1)!}}\ket{2n+1},
\end{aligned}
\end{equation} 
and with this choice of basis, $c_0 = e^{-\frac{\mu}{2}}\sqrt{\cosh(\mu)}$ and $c_1 = e^{-\frac{\mu}{2}}\sqrt{\sinh(\mu)}$. 

We then obtain a basis for span($\mathcal{S}$) as $\mathcal{B}=\{\ket{e_0,e_0},\ket{e_1,e_1},\ket{e_0, e_1},\ket{e_1,e_0}\}.$ We remark that this particular ordering of basis elements is useful for the presentation and later allows us to see the block diagonal structures of some particular POVM elements more straightforwardly. 

We now write out signal states in the set $\mathcal{S}$ as column vectors in this basis $\mathcal{B}$:
\begin{equation}\label{eq:statesVector}
\begin{aligned}
\ket{+\sqrt{\mu},+\sqrt{\mu}}&=\begin{pmatrix}c_0^2\\c_1^2\\c_0c_1\\c_0c_1\end{pmatrix}, &
\ket{-\sqrt{\mu},-\sqrt{\mu}}&=\begin{pmatrix}c_0^2\\c_1^2\\-c_0c_1\\-c_0c_1\end{pmatrix}, \\
\ket{+\sqrt{\mu},-\sqrt{\mu}}&=\begin{pmatrix}c_0^2\\-c_1^2\\-c_0c_1\\c_0c_1\end{pmatrix}, &
\ket{-\sqrt{\mu},+\sqrt{\mu}}&=\begin{pmatrix}c_0^2\\-c_1^2\\c_0c_1\\-c_0c_1\end{pmatrix}. \\
\end{aligned}
\end{equation}

Once we write out $F^{\gamma}$ in the basis $\mathcal{B}$, we can then find Eve's conditional states in the basis $\mathcal{B}$ by appropriate multiplications. Then the evaluation of von Neumann entropy of conditional states is straightforward since finding eigenvalues of $4 \times 4$ matrices is computationally simple.

\subsection{Change of basis matrix} 
Suppose we have determined $\bra{\vec{\alpha}}F^{\gamma}\ket{\vec{\beta}}$, where $\vec{\alpha}, \vec{\beta} \in \mathcal{S}$. Now we want to write out $F^{\gamma}$ in the basis $\mathcal{B}$. This can be done by a change of basis matrix.

We can write $\ket{\vec{e}_m} \in \mathcal{B}$ in the basis $\mathcal{S}$:
\begin{equation}
\ket{\vec{e}_m} =\sum_{n} A_{nm} \ket{\vec{\alpha}_n},
\end{equation}
where $A_{nm}$ is the $(n,m)$ entry of the desired change of basis matrix $A$ and $\ket{\vec{\alpha}_n} \in \mathcal{S}$. 

Similarly, 
\begin{equation}
\bra{\vec{e}_m} =\sum_{n} \bar{A}_{nm} \bra{\vec{\alpha}_n} = \sum_{n}{(A^{\dagger})_{mn}} \bra{\vec{\alpha}_n},
\end{equation}
where $\bar{A}_{nm}$ is the complex conjugate of $A_{nm}$ and $A^{\dagger}$ is the Hermitian conjugate of $A$.

Combining previous two equations, we have 
\begin{equation}\label{eq:Fentries}
\bra{\vec{e}_m}F^{\gamma}\ket{\vec{e}_n} = \sum_{i,j} (A^{\dagger})_{mj}\bra{\vec{\alpha}_j} F^{\gamma} \ket{\vec{\alpha}_i} A_{in}. 
\end{equation}

In the ordering of $\mathcal{S}$ and $\mathcal{B}$, this change of basis matrix $A$ can be expressed as
\begin{equation}
A=\begin{pmatrix}
\frac{1}{4c_0^2} & \frac{1}{4c_1^2} & \frac{1}{4c_0 c_1} & \frac{1}{4c_0 c_1} \\
\frac{1}{4c_0^2} & \frac{1}{4c_1^2} & -\frac{1}{4c_0 c_1} & -\frac{1}{4c_0 c_1} \\
\frac{1}{4c_0^2} & -\frac{1}{4c_1^2} & -\frac{1}{4c_0 c_1} & \frac{1}{4c_0 c_1} \\
\frac{1}{4c_0^2} & -\frac{1}{4c_1^2} & \frac{1}{4c_0 c_1} & -\frac{1}{4c_0 c_1}
\end{pmatrix},
\end{equation}
where $c_0$ and $c_1$ are defined from Eq. (\ref{eq:decomposition}).

\section{SIMULATION}\label{sec:simulationdetails}
 \begin{figure}
\includegraphics[scale=0.45]{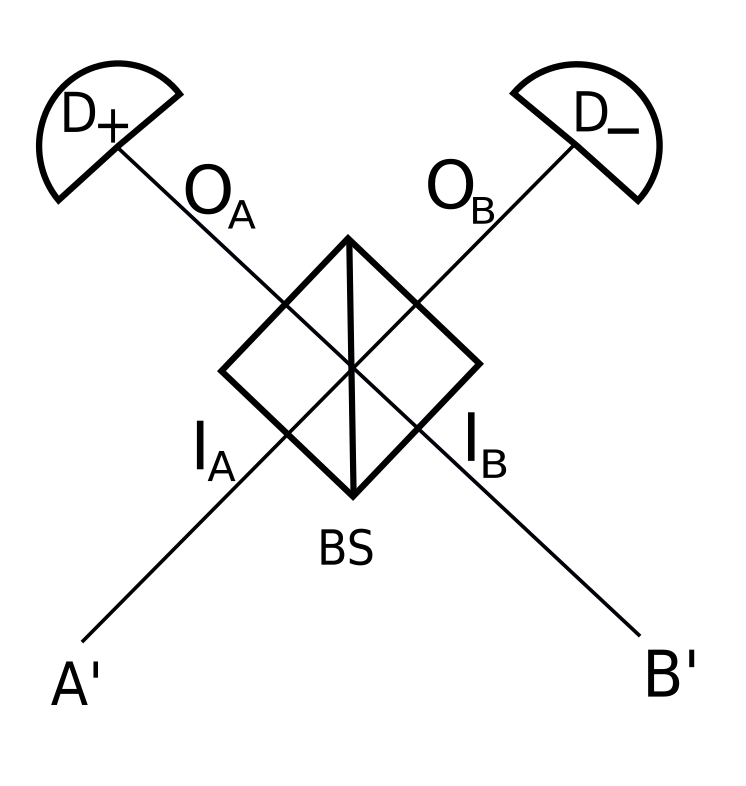}
 \caption{\label{fig:notation} Explanation of labels for input and output modes of the beam splitter. }
 \end{figure}
 
In this section, we explain how to obtain values of $\bra{\vec{\alpha}} F^{\gamma} \ket{\vec{\beta}}$ for $\ket{\vec{\alpha}}, \ket{\vec{\beta}} \in \mathcal{S}$ through simulations. After knowing  $\bra{\vec{\alpha}} F^{\gamma} \ket{\vec{\beta}}$, we can then use the results from Appendix \ref{sec:representation}, in particular, Eq. (\ref{eq:Fentries}), to express $F^{\gamma}$ in the basis $\mathcal{B}$ and then proceed with the evaluation of key rate.

To avoid the confusion between our simulation method and experimental execution of the protocol, we remark on how to obtain $\bra{\vec{\alpha}} F^{\gamma} \ket{\vec{\beta}}$ in the actual implementation of the protocol. In the parameter estimation step, the values of $\bra{\vec{\alpha}}F^{\gamma} \ket{\vec{\alpha}}$ for $\vec{\alpha} \in \mathcal{S}$ are directly obtained from observed correlation. The values of $\bra{\vec{\alpha}}F^{\gamma} \ket{\vec{\beta}}$ for $\vec{\alpha} \neq \vec{\beta}$ can be calculated by observed correlation of test states, as explained in Sec. \ref{sec:POVMDetermination}.

For our simulation, we propagate the input states through our simulated model of imperfections and then apply the POVM of detectors to the final states arriving at the detectors to calculate  $\bra{\vec{\alpha}} F^{\gamma} \ket{\vec{\beta}}$.

\subsection{Eve's POVM associated with loss-only scenario}\label{sec:simulationdetails:loss}
We now consider the loss-only scenario. Here, we simulate the quantum channel as a lossy channel and we consider the normal situation where Charlie (Eve) is honest and performs the measurements shown in Fig. \subref*{fig:scheme_a}. Namely, we calculate the POVM $F$ corresponding to the real setup with ideal devices at the central node. Our protocol can verify via test states in the test mode that this is the actual Eve's POVM in the loss-only scenario. As mentioned in the main text, for the purpose of presentation, we consider the symmetric setup. For a total single-photon transmissivity $\eta$ between Alice and Bob, each path has a transmissivity $\sqrt{\eta}$.

In this section, we will label Eve's POVM $F^{\gamma}$ associated with the loss-only scenario by adding the subscript ``loss". As shown in Fig. (\ref{fig:notation}), we will label the input modes of the beam splitter at the central node as $I_A$ and $I_B$ and the output modes as $O_A$ and $O_B$, where $O_A$ reaches the detector $D_+$ and $O_B$ reaches the detector $D_-$.

We describe how to obtain $\bra{\vec{\alpha}} F_{\text{loss}}^{\gamma} \ket{\vec{\beta}}$. First, Alice and Bob prepare coherent states $\ket{\alpha_A}_{A'}$ and $\ket{\alpha_B}_{B'}$ in the registers $A'$ and $B'$, respectively, and send them to Charlie. After the lossy channel, the state becomes $\ket{\sqrt{\sqrt{\eta}}\alpha_A, \sqrt{\sqrt{\eta}}\alpha_B}_{I_AI_B}$, and Eve has the state $\ket{\sqrt{1-\sqrt{\eta}}\alpha_A, \sqrt{1-\sqrt{\eta}}\alpha_B}_{E_AE_B}$ at her disposal.  Then after the beam splitter, the state becomes $\ket{\frac{\sqrt{\sqrt{\eta}}\alpha_A+ \sqrt{\sqrt{\eta}}\alpha_B}{\sqrt{2}},\frac{\sqrt{\sqrt{\eta}}\alpha_A- \sqrt{\sqrt{\eta}}\alpha_B}{\sqrt{2}}}_{O_AO_B}$. We now apply the POVM of the detectors to this state.

The ideal detectors are described by the following POVM:
\begin{equation}\label{eq:detectorPOVM}
\begin{aligned}
\Pi_{\text{ideal}}^{+} &= (\mathds{1}_{O_A} -\dyad{0}{0}_{O_A}) \otimes \dyad{0}{0}_{O_B},\\
\Pi_{\text{ideal}}^{-} & = \dyad{0}{0}_{O_A} \otimes (\mathds{1}_{O_B} -\dyad{0}{0}_{O_B}),\\
\Pi_{\text{ideal}}^{?} &= \dyad{0}{0}_{O_A} \otimes \dyad{0}{0}_{O_B},\\
 \Pi_{\text{ideal}}^{d} & = (\mathds{1}_{O_A}-\dyad{0}{0}_{O_A}) \otimes (\mathds{1}_{O_B}-\dyad{0}{0}_{O_B}),
\end{aligned}
\end{equation}
where $\mathds{1}$ is the identity operator and $\ket{0}$ is the vacuum state.

Then, for $\ket{\vec{\alpha}} = \ket{\alpha_A,\alpha_B}$ and $\ket{\vec{\beta}}=\ket{\beta_A,\beta_B}$, we evaluate $\bra{\vec{\alpha}}F_{\text{loss}}^{\gamma} \ket{\vec{\beta}}$ as
\begin{widetext}
\begin{equation}\label{eq:FlossSbasis}
\begin{aligned}
\bra{\vec{\alpha}}F_{\text{loss}}^{\gamma} \ket{\vec{\beta}} =& \bra{\frac{\eta^{\frac{1}{4}}(\alpha_A+ \alpha_B)}{\sqrt{2}},\frac{\eta^{\frac{1}{4}}(\alpha_A- \alpha_B)}{\sqrt{2}}}\Pi_{\text{ideal}}^{\gamma} \ket{\frac{\eta^{\frac{1}{4}}(\beta_A+ \beta_B)}{\sqrt{2}},\frac{\eta^{\frac{1}{4}}(\beta_A- \beta_B)}{\sqrt{2}}}\braket{\sqrt{1-\sqrt{\eta}}\alpha_A}{\sqrt{1-\sqrt{\eta}}\beta_A}_{E_A} \\ &\times \braket{\sqrt{1-\sqrt{\eta}}\alpha_B}{\sqrt{1-\sqrt{\eta}}\beta_B}_{E_B}.
\end{aligned}
\end{equation}
\end{widetext}

Now we have obtained values for $\bra{\vec{\alpha}}F_{\text{loss}}^{\gamma}\ket{\vec{\beta}}$ from simulations, and we can then proceed to write $F_{\text{loss}}^{\gamma}$ in the basis $\mathcal{B}$ by using Eq. (\ref{eq:Fentries}).
\begin{equation}\label{eq:Floss}
\begin{aligned}
F_{\text{loss}}^{+} &= (1-\xi^2) \begin{pmatrix}
\frac{1- \xi^2 \Omega^2}{8c_0^4} & \frac{1- \xi^2 \Omega^2}{8c_0^2 c_1^2} & 0 & 0\\
\frac{1- \xi^2 \Omega^2}{8c_0^2 c_1^2} & \frac{1- \xi^2 \Omega^2}{8c_1^4} & 0 & 0\\
0&0 &\frac{1+\xi^2 \Omega^2}{8c_0^2 c_1^2} & \frac{1+ \xi^2 \Omega^2}{8c_0^2 c_1^2} \\
0&0 &\frac{1+ \xi^2 \Omega^2}{8c_0^2 c_1^2} & \frac{1+\xi^2 \Omega^2}{8c_0^2 c_1^2} 
\end{pmatrix},\\
F_{\text{loss}}^{-} &= (1-\xi^2) \begin{pmatrix}
\frac{1- \xi^2 \Omega^2}{8c_0^4} & \frac{-1+\xi^2 \Omega^2}{8c_0^2 c_1^2} & 0 & 0\\
\frac{ -1+\xi^2 \Omega^2}{8c_0^2 c_1^2} & \frac{1- \xi^2 \Omega^2}{8c_1^4} & 0 & 0\\
0&0 &\frac{1+\xi^2 \Omega^2}{8c_0^2 c_1^2} & \frac{-1-\xi^2 \Omega^2}{8c_0^2 c_1^2} \\
0&0 &\frac{-1- \xi^2 \Omega^2}{8c_0^2 c_1^2} & \frac{1+\xi^2 \Omega^2}{8c_0^2 c_1^2} 
\end{pmatrix}, \\
F_{\text{loss}}^{?} &= \xi^2 \begin{pmatrix}
\frac{(1+\Omega)^2}{4c_0^4} &0 & 0 & 0\\
0 & \frac{(1-\Omega)^2}{4c_1^4} & 0 & 0\\
0&0 &\frac{1-\Omega^2}{4c_0^2 c_1^2} &0 \\
0&0 &0 & \frac{1-\Omega^2}{4c_0^2 c_1^2} 
\end{pmatrix},\\
F_{\text{loss}}^d &= 0,
\end{aligned}
\end{equation}
where for the ease of representation, we define $\Omega = e^{-2(1-\sqrt{\eta})\mu}$, and $\xi = e^{- \sqrt{\eta} \mu}$.  Also, $c_0$ and $c_1$ are defined from the decomposition in Eq. (\ref{eq:decomposition}). By noting that $2c_0^2 = 1 + \xi^2 \Omega$ and $2c_1^2 =  1 - \xi^2 \Omega$, we can easily check that $F_{\text{loss}}^+ +F_{\text{loss}}^- + F_{\text{loss}}^? + F_{\text{loss}}^d = \mathds{1}.$ 

\subsection{Models for imperfections}\label{sec:simulationdetails:models}
In this section, we consider realistic imperfections in the experimental setup. Specifically, we consider the mode mismatch, phase mismatch, dark counts of detectors and the inefficiency of detectors. In the following, we describe the physical models for those imperfections. 

\subsubsection{Mode mismatch}

We consider the mode mismatch with a simulation parameter $V$. The model of the mode mismatch is that for a coherent state $\ket{\alpha}_1$ in a mode $1$, due to the mode mismatch, the state $\ket{\sqrt{V}\alpha}_1$ remains in the mode $1$ and $\ket{\sqrt{1-V}\alpha}_2$ is in the mode $2$, which is distinct from the mode $1$. To derive values for $\bra{\vec{\alpha}}F^{\gamma} \ket{\vec{\beta}}$, we will propagate the input states, similar to the loss-only case. In the place of $I_A, I_B, O_A, O_B$ used in the discussion of loss-only scenario, we will replace them by $I_{A1}, I_{B1}, O_{A1}, O_{B1}$ for the initial mode and $I_{A2}, I_{B2}, O_{A2}, O_{B2}$ for the additional mode.

Suppose Alice sends $\ket{\alpha_A}_{A'}$ and Bob sends $\ket{\alpha_B}_{B'}$. Since only the relative mode mismatch between Alice's mode and Bob's mode matters, without loss of generality, we leave the state in the register $A'$ untouched when it reaches $I_{A1}$, that is, we have $\ket{\alpha_A}_{I_{A1}}$. Due to the mode mismatch, the state $\ket{\alpha_B}_{B'}$ becomes $\ket{\sqrt{V}\alpha_B}_{I_{B1}}\ket{\sqrt{1-V}\alpha_B}_{I_{B2}}$. Correspondingly, we have the vacuum state in the mode $I_{A2}$. To summarize, the state arriving at the beam splitter of Charlie's station due to mode mismatch is $\ket{\alpha_A,0}_{I_{A1}I_{A2}}\ket{\sqrt{V}\alpha_B,\sqrt{1-V}\alpha_B}_{I_{B1}I_{B2}}$. The mode $I_{A1}$ interferes with the mode $I_{B1}$ and the mode $I_{A2}$ interferes with the mode $I_{B2}$ independently. We remark here that this parameter $V$ can be made close to 1 with experimentally available compensation systems, for example, see Ref. \cite{Liu2013} in the setting of MDI protocols. In particular, $V\geq 95\%$ is readily achievable. 

\subsubsection{Phase mismatch}

Ideally, Alice and Bob should prepare coherent states with the same global phase. We consider the situation where there is a phase mismatch between Alice's signal state and Bob's signal state. For the ideal input state $\ket{\alpha_A,\alpha_B}_{A'B'}$, due to the phase mismatch, the state becomes $\ket{\alpha_A,\alpha_B e^{i \delta}}_{I_AI_B}$ for some $\delta$.  We expect that with an experimentally feasible phase compensation system, the value of $\delta$ is typically small. For instance, the continuous-variable QKD experiment in Ref. \cite{Soh2015} reports a value less than $\frac{\pi}{60}$.

\subsubsection{Detector dark count}
We now consider dark counts of detectors. For simplicity of our presentation, we model two detectors to have the same dark count probability $p_d$. It is also straightforward to model the case where two detectors have different dark count probabilities. 

The effect of dark counts can be taken into consideration by modifying the POVM for detectors. Eq. (\ref{eq:detectorPOVM}) gives the POVM associated with ideal detectors. When the detectors have dark counts, the associated POVM is modified as below
\begin{equation}
\begin{aligned}
\Pi_{\text{dark}}^{+} =& (\mathds{1}_{O_A} -\dyad{0}{0}_{O_A})  \otimes (1-p_d) \dyad{0}{0}_{O_B} \\
&+ p_d \dyad{0}{0}_{O_A} \otimes (1-p_d) \dyad{0}{0}_{O_B}
\\ =& (1-p_d) \Pi_{\text{ideal}}^{+} + (1-p_d)p_d \Pi_{\text{ideal}}^{?}, \\
\Pi_{\text{dark}}^{-}  =& (1-p_d) \dyad{0}{0}_{O_A}  \otimes (\mathds{1}_{O_B}  -\dyad{0}{0}_{O_B} )\\
&+  (1-p_d) \dyad{0}{0}_{O_A} \otimes p_d \dyad{0}{0}_{O_B} \\
=&  (1-p_d) \Pi_{\text{ideal}}^{-} + (1-p_d)p_d \Pi_{\text{ideal}}^{?},  \\ 
\Pi_{\text{dark}}^{?} =& (1-p_d)\dyad{0}{0}_{O_A} \otimes (1-p_d)\dyad{0}{0}_{O_B} \\
 =& (1-p_d)^2 \Pi_{\text{ideal}}^{?}, \\
\Pi_{\text{dark}}^{d}  =&   (\mathds{1}_{O_A} -\dyad{0}{0}_{O_A}) \otimes p_d\dyad{0}{0}_{O_B}  \\
& +  p_d \dyad{0}{0}_{O_A} \otimes (\mathds{1}_{O_B} -\dyad{0}{0}_{O_B} ) \\
& + p_d \dyad{0}{0}_{O_A} \otimes p_d\dyad{0}{0}_{O_B}  \\
& + (\mathds{1}_{O_A} -\dyad{0}{0}_{O_A}) \otimes (\mathds{1}_{O_B} -\dyad{0}{0}_{O_B} )  \\
 =&  p_d \Pi_{\text{ideal}}^{+} + p_d \Pi_{\text{ideal}}^{-} + p_d^2 \Pi_{\text{ideal}}^{?}+ \Pi_{\text{ideal}}^{d}.
\end{aligned}
\end{equation}

Since in our simulation, we propagate the input states through the physical models of imperfections to derive the final states before the detectors and then use the POVM of detectors to derive the values of $\bra{\vec{\alpha}}F^{\gamma}\ket{\vec{\beta}}$, the expression of $\bra{\vec{\alpha}}F^{\gamma}\ket{\vec{\beta}}$ will have a similar structure as the loss-only case shown in Eq. (\ref{eq:FlossSbasis}). From this observation, we know that once we obtained the expression of Eve's POVM element $F^{\gamma}$, when we have considered all other imperfections except dark counts, we can then derive the POVM elements including dark counts of detectors by probabilistic mixtures of $F^{\gamma}$, following the same relation as between $\Pi_{\text{dark}}^{\gamma}$ and $\Pi_{\text{ideal}}^{\gamma}$. 

To illustrate the idea, we give a simple example where we consider the physical channel to be a lossy channel and we want to include dark counts of detectors in our simulation. Since we have derived $F^{\gamma}_{\text{loss}}$ in Eq. (\ref{eq:Floss}), we can derive the expressions of $F^{\gamma}_{\text{dark}}$ corresponding to this simulation as follows:
\begin{equation}
\begin{aligned}
F_{\text{dark}}^{+} & = (1-p_d) F_{\text{loss}}^{+} + (1-p_d)p_d F_{\text{loss}}^{?}, \\
F_{\text{dark}}^{-} & = (1-p_d) F_{\text{loss}}^{-} + (1-p_d)p_d F_{\text{loss}}^{?},  \\
F_{\text{dark}}^{?} & = (1-p_d)^2 F_{\text{loss}}^{?}, \\
F_{\text{dark}}^{d} & =   p_d F_{\text{loss}}^{+} + p_d F_{\text{loss}}^{-} + p_d^2 F_{\text{loss}}^{?} +F_{\text{loss}}^{d}.
\end{aligned}
\end{equation}

\subsubsection{Detector's efficiency}
We take into account that any practical single-photon detector has a limited efficiency. In our simulation method, we can easily modify the POVM of detectors as in Eq. (\ref{eq:detectorPOVM}), to include the efficiency of each detector separately. However, for simplicity of our presentation, we assume that both detectors have the same efficiency $\eta_d$ so that we can combine the detector's efficiency and the channel transmittance by redefining the total transmissivity $\eta$. Let $\eta_t$ refer to the single-photon transmission probability of the quantum channel between Alice and Bob. Since both detectors have the same efficiency, we can redefine the total transmissivity $\eta = \eta_t \eta_d^2$ and use this value of $\eta$ in the simulation. Then we can still use the POVM of detectors with the perfect efficiency in our simulation.

\subsection{Eve's POVM with those imperfections}\label{sec:simulationdetails:POVM}
As discussed in the previous section, we now take into consideration the mode mismatch with a simulation parameter $V$, and the phase mismatch with a simulation parameter $\delta$. We consider both detectors have the same detector efficiency $\eta_d$ and the same dark count probability $p_d$.

We will first derive Eve's POVM $F^{\gamma}_{\text{mismatch}}$ when we consider both the mode mismatch and the phase mismatch. Then we derive Eve's POVM $F^{\gamma}_{\text{model}}$ when we include dark counts of detectors as well. Finally, the effects of detector efficiency is taken into consideration by a redefinition of $\eta$.

For an input coherent state $\ket{\alpha_A, \alpha_B}_{A'B'}$, the state after the lossy channels and models for the mode mismatch and phase mismatch becomes $\ket{\sqrt{\sqrt{\eta}}\alpha_A, \sqrt{\sqrt{\eta}} \sqrt{V}\alpha_B e^{i\delta}}_{I_{A1}I_{B1}} \otimes \ket{0, \sqrt{\sqrt{\eta}} \sqrt{1-V}\alpha_B e^{i\delta}}_{I_{A2} I_{B2}}$ and Eve has  $\ket{\sqrt{1-\sqrt{\eta}}\alpha_A, \sqrt{1-\sqrt{\eta}}\alpha_B}_{E_AE_B}$ at her disposal. Then, the state arriving at the detectors is $\ket{\frac{\sqrt{\sqrt{\eta}}\alpha_A+ \sqrt{\sqrt{\eta}}\sqrt{V}\alpha_Be^{i\delta}}{\sqrt{2}},\frac{\sqrt{\sqrt{\eta}}\alpha_A- \sqrt{\sqrt{\eta}}\sqrt{V}\alpha_Be^{i\delta}}{\sqrt{2}}}_{O_{A1}O_{B1}} \otimes \ket{\frac{ \sqrt{\sqrt{\eta}}\sqrt{1-V}\alpha_Be^{i\delta}}{\sqrt{2}},-\frac{ \sqrt{\sqrt{\eta}}\sqrt{1-V}\alpha_Be^{i\delta}}{\sqrt{2}}}_{O_{A2}O_{B2}}$.

We now introduce the POVM of the ideal detectors when there are two independent modes entering the detectors due to mode mismatch. 

 \begin{equation}\label{eq:mismatchdetectorPOVM}
\begin{aligned}
\Pi_{\text{mismatch}}^{+} =& \ (\mathds{1}_{O_{A1}O_{A2}} -\dyad{00}{00}_{O_{A1}O_{A2}}) \otimes \dyad{00}{00}_{O_{B1}O_{B2}},\\
\Pi_{\text{mismatch}}^{-} =& \ \dyad{00}{00}_{O_{A1}O_{A2}} \otimes (\mathds{1}_{O_{B1}O_{B2}} -\dyad{00}{00}_{O_{B1}O_{B2}}),\\
\Pi_{\text{mismatch}}^{?} =& \ \dyad{00}{00}_{O_{A1}O_{A2}} \otimes \dyad{00}{00}_{O_{B1}O_{B2}},\\
 \Pi_{\text{mismatch}}^{d} =& \ (\mathds{1}_{O_{A1}O_{A2}}-\dyad{00}{00}_{O_{A1}O_{A2}})\\
 &  \otimes (\mathds{1}_{O_{B1}O_{B2}}-\dyad{00}{00}_{O_{B1}O_{B2}}).
\end{aligned}
\end{equation}

Then, for $\ket{\vec{\alpha}} = \ket{\alpha_A,\alpha_B}$ and $\ket{\vec{\beta}}=\ket{\beta_A,\beta_B}$, we first define
\begin{widetext}
\begin{equation}
\begin{aligned}
\ket{\tilde{\alpha}_{\text{final}}} &=   \ket{\frac{\eta^{1/4}(\alpha_A+\sqrt{V}\alpha_Be^{i\delta})}{\sqrt{2}}, \frac{\eta^{1/4}\sqrt{1-V}\alpha_Be^{i\delta}}{\sqrt{2}},\frac{\eta^{1/4}(\alpha_A- \sqrt{V}\alpha_Be^{i\delta})}{\sqrt{2}},-\frac{\eta^{1/4} \sqrt{1-V}\alpha_Be^{i\delta}}{\sqrt{2}}}_{O_{A1}O_{A2}O_{B1}O_{B2}}, \\
\ket{\tilde{\beta}_{\text{final}}} &=   \ket{\frac{\eta^{1/4}(\beta_A+\sqrt{V}\beta_Be^{i\delta})}{\sqrt{2}}, \frac{\eta^{1/4}\sqrt{1-V}\beta_Be^{i\delta}}{\sqrt{2}},\frac{\eta^{1/4}(\beta_A- \sqrt{V}\beta_Be^{i\delta})}{\sqrt{2}},-\frac{\eta^{1/4} \sqrt{1-V}\beta_Be^{i\delta}}{\sqrt{2}}}_{O_{A1}O_{A2}O_{B1}O_{B2}}.
\end{aligned}
\end{equation}

We evaluate $\bra{\vec{\alpha}}F_{\text{mismatch}}^{\gamma} \ket{\vec{\beta}}$ as

\begin{equation}
\begin{aligned}
\bra{\vec{\alpha}}F_{\text{mismatch}}^{\gamma} \ket{\vec{\beta}} &=
\bra{\tilde{\alpha}_{\text{final}}} \Pi^{\gamma}_{\text{mismatch}} \ket{\tilde{\beta}_{\text{final}}}\braket{\sqrt{1-\sqrt{\eta}}\alpha_A}{\sqrt{1-\sqrt{\eta}}\beta_A}_{E_A} \braket{\sqrt{1-\sqrt{\eta}}\alpha_B}{\sqrt{1-\sqrt{\eta}}\beta_B}_{E_B},
\end{aligned}
\end{equation}
where $\eta = \eta_t \eta_d^2$ and $\eta_t = 10^{-\frac{0.2L}{10}}$ for a distance $L$ in km.

Now, we write down $F^{\gamma}_{\text{mismatch}}$ in the basis $\mathcal{B}$. For the ease of representation, we define $\xi = e^{-\sqrt{\eta} \mu}$ and $\Omega = e^{-2(1-\sqrt{\eta})\mu}$ as before. We then define
\begin{equation}
\begin{aligned}
 a&=(1-\xi^{(1+\sqrt{V}\cos{\delta})})\xi^{(1-\sqrt{V}\cos{\delta})}, \\ b&=(\xi^{2(1+\sqrt{V}\cos{\delta})}-\xi^{(1+\sqrt{V}\cos{\delta})})\xi^{(1-\sqrt{V}\cos{\delta})}\Omega^2, \\ 
 c&=(\xi^{1+i\sqrt{V}\sin{\delta}} -\xi)\xi \Omega,\\ 
  d&=(\xi^{1-i\sqrt{V}\sin{\delta}} -\xi)\xi \Omega, \\  
  o&=(1-\xi^{(1-\sqrt{V}\cos{\delta})})\xi^{(1+\sqrt{V}\cos{\delta})},\\
   p&=(\xi^{2(1-\sqrt{V}\cos{\delta})}-\xi^{(1-\sqrt{V}\cos{\delta})})\xi^{(1+\sqrt{V}\cos{\delta})}\Omega^2, \\  
   q&= (\xi^{1+i\sqrt{V}\sin{\delta}} -\xi) (\xi^{1-i\sqrt{V}\sin{\delta}}-\xi)\Omega,\\
m&=(1-\xi^{(1+\sqrt{V}\cos{\delta})})(1-\xi^{(1-\sqrt{V}\cos{\delta})}),\\  
n&=(\xi^{2(1+\sqrt{V}\cos{\delta})}-\xi^{(1+\sqrt{V}\cos{\delta})})(\xi^{2(1-\sqrt{V}\cos{\delta})}-\xi^{(1-\sqrt{V}\cos{\delta})})\Omega^2.
 \end{aligned}
 \end{equation} 
 \end{widetext}
 \onecolumngrid
 And, $c_0$ and $c_1$ are again defined from the decomposition in Eq. (\ref{eq:decomposition}). Then we have
\begin{equation}
\begin{aligned}
F_{\text{mismatch}}^{+} &=  \begin{pmatrix}
\frac{a+b+2c+2d+o+p}{8c_0^4} & \frac{a+b-o-p}{8c_0^2 c_1^2} & 0 & 0\\
\frac{a+b-o-p}{8c_0^2 c_1^2} & \frac{a+b-2c-2d+o+p}{8c_1^4} & 0 & 0\\
0&0 &\frac{a-b+o-p}{8c_0^2 c_1^2} & \frac{a-b+2c-2d-o+p}{8c_0^2 c_1^2} \\
0&0 &\frac{a-b-2c+2d-o+p}{8c_0^2 c_1^2} & \frac{a-b+o-p}{8c_0^2 c_1^2} 
\end{pmatrix},\\
F_{\text{mismatch}}^{-} &=  \begin{pmatrix}
\frac{a+b+2c+2d+o+p}{8c_0^4} & -\frac{a+b-o-p}{8c_0^2 c_1^2} & 0 & 0\\
-\frac{a+b-o-p}{8c_0^2 c_1^2} & \frac{a+b-2c-2d+o+p}{8c_1^4} & 0 & 0\\
0&0 &\frac{a-b+o-p}{8c_0^2 c_1^2} & -\frac{a-b+2c-2d-o+p}{8c_0^2 c_1^2} \\
0&0 &-\frac{a-b-2c+2d-o+p}{8c_0^2 c_1^2} & \frac{a-b+o-p}{8c_0^2 c_1^2} 
\end{pmatrix}, \\
F_{\text{mismatch}}^{?} &=\xi^2 \begin{pmatrix}
\frac{(1+\Omega)^2}{4c_0^4} &0 & 0 & 0\\
0 & \frac{(1-\Omega)^2}{4c_1^4} & 0 & 0\\
0&0 &\frac{1-\Omega^2}{4c_0^2 c_1^2} &0 \\
0&0 &0 & \frac{1-\Omega^2}{4c_0^2 c_1^2} 
\end{pmatrix},\\
F_{\text{mismatch}}^d &= \begin{pmatrix}
\frac{m+n+2q}{4c_0^4} &0 & 0 & 0\\
0 & \frac{m+n-2q}{4c_1^4}  & 0 & 0\\
0&0 &\frac{m-n}{4c_0^2 c_1^2} &0 \\
0&0 &0 & \frac{m-n}{4c_0^2 c_1^2} 
\end{pmatrix}.
\end{aligned}
\end{equation}

Finally, Eve's effective POVM corresponding to the mode mismatch, phase mismatch and dark counts of detectors is given as below
\begin{equation}\label{eq:simulationPOVM}
\begin{aligned}
F_{\text{model}}^{+} & = (1-p_d) F_{\text{mismatch}}^{+} + (1-p_d)p_d F_{\text{mismatch}}^{?}, \; \; F_{\text{model}}^{-}  = (1-p_d) F_{\text{mismatch}}^{-} + (1-p_d)p_d F_{\text{mismatch}}^{?}, \\
F_{\text{model}}^{?} & = (1-p_d)^2 F_{\text{mismatch}}^{?}, \; \; F_{\text{model}}^{d} =  p_d F_{\text{mismatch}}^{+} + p_d F_{\text{mismatch}}^{-}+p_d^2 F_{\text{mismatch}}^{?}  + F_{\text{mismatch}}^{d}.
\end{aligned}
\end{equation}

\twocolumngrid

\bibliographystyle{apsrev4-1.bst}
\bibliography{PMMDIQKD}

\end{document}